\documentclass{aastex6}

\newcommand{\kms}{km~s{$^{-1}$}}
\newcommand{\msol}{M{$_{\odot}$}}
\newcommand{\Msol}{M{$_{\odot}$}}

\newcommand{\Lsol}{L{$_{\odot}$}}
\newcommand{\Vlsr}{{V$_{LSR}$}}

\newcommand{\Tco}{{$^{12}$CO}}
\newcommand{\co}{C{$^{18}$O}}
\newcommand{\Feii}{Fe~{\sc ii}}

\newcommand{\Htwo}{H{$_2$}}
\newcommand{\per}{$^{-1}$}

\begin{document}


\title{The ALMA View of the OMC1 Explosion in Orion}

%% Use \author, \affil, plus the \and command to format author and affiliation 
%% information.  If done correctly the peer review system will be able to
%% automatically put the author and affiliation information from the manuscript
%% and save the corresponding author the trouble of entering it by hand.
%%
%% The \affil should be used to document primary affiliations and the
%% \altaffil should be used for secondary affiliations, titles, or email.

%% Authors with the same affiliation can be grouped in a single

\author{John Bally\altaffilmark{1}}
\affil{Astrophysical and Planetary Sciences Department \\
       University of Colorado, UCB 389 Boulder, Colorado 80309, USA}
       
\author{Adam Ginsburg\altaffilmark{2}}
% \affil{ ESO Headquarters \\
%       Karl-Schwarzschild-Str. 2 85748, Garching bei Munchen, Germany}
\affil{ Jansky Fellow, National Radio Astronomy Observatory \\
        1003 Lopezville Rd., Socorro, NM 87801, USA }

\author{Hector Arce}
\affil{Department of Astronomy  \\
        Steinbach Hall, 52 Hillhouse Ave., Yale University, New Haven, CT 06511, USA}

\author{Josh Eisner}
\affil{Steward Observatory \\
        University of Arizona, 933 North Cherry Avenue, Tucson, AZ 85721, USA}

\author{Allison Youngblood}
\affil{Astrophysical and Planetary Sciences Department \\
       University of Colorado, UCB 389 Boulder, Colorado 80309, USA}

\author{Luis Zapata}
\affil{Instituto de Radioastronom\'ia y Astrof\'i's\'ica, UNAM \\
       Apdo. Postal 3-72 (Xangari), 58089 Morelia, Michoac‡\'an, M\'exico }
       
\and

\author{Hans Zinnecker}
\affil{Deutsches SOFIA Institut (DSI) \\
        University of Stuttgart, Pfaffenwaldring 29, 70569, Germany}

\altaffiltext{1}{john.bally@colorado.edu}

\altaffiltext{2}{Adam Ginsburg is a Jansky Fellow of the National Radio Astronomy Observatory.}

\begin{abstract}

Most massive stars form in dense clusters where gravitational interactions 
with other stars may be common.     The two nearest forming massive stars, 
the BN object and Source I,  located behind the Orion Nebula,  were 
ejected with velocities  of  $\sim$29  and $\sim$13 \kms\  about 500 years  
ago by such interactions.  This event generated an explosion in the gas.   
New ALMA  observations  show  in unprecedented detail, a roughly 
spherically symmetric  distribution of over a hundred  \Tco\ J=2$-$1  
streamers  with velocities  extending from  \Vlsr =$-$150 to +145 \kms. 
The streamer radial velocities increase (or decrease) linearly with 
projected distance from the explosion center,  forming a  `Hubble Flow'
confined to within 50\arcsec\ of the explosion center.    They point toward 
the high proper-motion,  shock-excited  \Htwo\ and [\Feii ]   `fingertips'  
and lower-velocity  CO  in  the \Htwo\ wakes  comprising  Orion's `fingers'.
In some directions, the \Htwo\ `fingers'  extend more than a factor of two 
farther from the ejection center  than the CO streamers.     
Such deviations from spherical symmetry 
may be caused  by ejecta running into dense gas or the dynamics of the 
N-body interaction  that ejected the stars and produced the explosion.   
This $\sim$10$^{48}$ erg event may have been powered by the release  
of gravitational potential energy  associated with the formation of a compact 
binary or a protostellar merger.    Orion may be the prototype   for a new  
class  of stellar  explosion   responsible  for luminous  infrared   transients  
in nearby galaxies.  

\end{abstract}




\section{Introduction} 

Quasi-steady accretion from a circumstellar accretion disk onto forming young stellar objects (YSOs) tends to power long-lived, collimated bipolar outflows and jets common in star forming regions \citep{Frank2014,BallyReipurth2001,Bally2016}.   Short-lived, powerful explosions,  accompanied by luminous infrared flares can be powered by gravitational interactions of 3 or more stars in dense star clusters that can lead to the formation of binaries by capture and stellar mergers  \citep{BallyZinnecker2005,SokerTylenda2006,MoeckelBally2007a,MoeckelBally2007b,PortegiesZwart2016}.   Such an explosion occurred behind the Orion Nebula $\sim$500 years ago  \citep{Zapata2009c,Bally2011,BallyGinsburg2015}. 

\subsection{Ejection of Runaway Stars:  BN, Source I, and Source  n}

The $\sim$100 \Msol\  Orion Molecular Core 1 (OMC1) located behind the Nebula \citep{GenzelStutzki1989} at a distance of $\sim$414 pc \citep{Menten2007,Goddi2011b}  is the densest and most molecule-rich part of the degree-long 2,200 \Msol\ Integral-Shaped Filament \citep{JohnstoneBally1999} in the   50,000 \msol\  Orion A cloud \citep{Bally1987}.  OMC1  contains the $\sim 1-2 \times 10^4$ \Lsol\  but visually obscured Becklin-Neugebauer (BN) object at the north-end of the Kleinmann-Low (KL)  infrared nebula  $\sim$1\arcmin\ northwest of the Trapezium  stars that  ionize the nebula.   BN, thought to be an 8  to 15 \Msol\ star \citep{Scoville1983,Tan2004,Goddi2011,ChatterjeeTan2012},  is moving towards the northwest with respect to other stars in the Nebula.  The proper motion has been measured to be 
21.3 \kms\  towards PA = 331\arcdeg\ \citep{Gomez2008},
26.4 \kms\  towards PA = 341\arcdeg\ \citep{Goddi2011}, and 
26.6 \kms\  towards PA = 323\arcdeg\ \citep{Dzib2016,Rodriguez2016}.   BN has a redshifted radial velocity of $\sim$ +11 \kms\ with respect to the \Vlsr\ = +9 \kms\  radial velocity of OMC1 \citep{Gomez2008,Goddi2011,Scoville1983}.  

It has been proposed that the BN object was ejected by  $\theta ^1$Ori C, the most massive ($\sim$40 \Msol ) member of  the Trapezium about 4,000 years ago \citep{Tan2004,ChatterjeeTan2012}.    However, another massive star embedded in OMC1, the highly-obscured radio-source, Source I, located $\sim$10" southeast of BN \citep{Menten1995},  is moving towards the southeast  in a  direction nearly opposite to the motion of  BN.   The proper motion of Source I has been measured to be 
14.4 \kms\ towards PA = 142\arcdeg\ \citep{Gomez2008}, 
11.5 \kms\  towards PA = 109\arcdeg\ \citep{Goddi2011}, and 
9.6  \kms\  towards PA = 159\arcdeg\ \citep{Dzib2016,Rodriguez2016}.  Source I  has a blueshifted radial velocity of $\sim$ $-$4 \kms\ with respect to OMC1, and is thus moving in a direction roughly opposite to BN  \citep{Plambeck2009,Plambeck2016}.  

The trajectories of BN and radio Source I  intersected within a 1" ($\sim$400 AU) diameter region between their current locations  about  500 years ago, indicating that they may have been ejected from there \citep{Gomez2008,Goddi2011}.    If this model is correct,  momentum conservation implies that Source I  must have a mass of $\sim$22$\pm$3 \Msol \citep{Goddi2011}.   However, mass estimates based on the rotation curve of the Source I disk suggest a considerably lower mass between 5 to 7 \Msol\  \citep{Plambeck2016,Plambeck2013,Matthews2010}.   
   
Another star known as infrared Source n \citep{Lonsdale1982}, located a few arc-seconds southwest of Source I,  was also found to be moving with a proper motion of $\sim$26 \kms\  towards  PA = 180$^o$ away from the region from which Source I and BN were ejected   \citep{Gomez2008}.    However,  \citet{Goddi2011}  failed to detect any motion.    Recently, the radio proper motion of Source n was re-measured and found to have a value of $\sim$15  \kms\ towards PA = 179$^o$ \citep{Rodriguez2016}.  Source n, which  in the radio is a $\sim$0.4\arcsec\ separation double source,   was found to have ejected a one-sided radio jet around 2006 which has affected previous determinations of its proper motion  \citep{Rodriguez2016}.    The luminosity of Source n is poorly constrained to be around 2,000 \Lsol\ implying a mass of  around 3 to 6 \Msol\ \citep{Greenhill2004}.     The errors on the most recent proper motion measurements of BN, Source I, and n by \citet{Dzib2016,Rodriguez2016} are about 2 \kms\ and $\pm$ 1\arcdeg\ to 2\arcdeg\  in total velocity on the plane of the sky and in position angle.     For Source I and BN, the variations in the measured proper motions mostly reflect different choices of the reference frame in which the motions are measured.   Thus, the OMC1 cloud core appears to have experienced a dynamical decay of a system of massive stars.  
 
Infrared source IRc4, located $\sim$6"  southwest of Source I was also found to be a $\sim 2 \times 10^4$ \Lsol\ protostar \citep{deBuzier2012}.    Thus, in addition to the ejected sources, there may be additional massive protostars embedded in OMC1.    Radio Source I, the BN object, and IRc4 are the most luminous objects  in OMC1 which has a luminosity of $\sim$10$^5$ \Lsol .  

 There are two models for the ejection of runaway stars \citep{GiesBolton1986}.  When the most massive member of a binary explodes as a supernova, the orbital motion of the companion can be converted into linear motion \citep{Tauris1998,Hoogerwerf2000,Hoogerwerf2001,Dray2005}.  Alternatively, ejections can occur by N-body interactions in dense clusters, interactions between single stars and binaries, binaries and binaries, or the dynamical decay of non-hierarchical groups \citep{Gualandris2004,Gvaramadze2011}.   

The acceleration of young runaway stars requires the  dynamical interaction of at least 3 stars which leads to the formation or hardening of a compact binary whose gravitational binding energy powers the ejection \citep{Reipurth2010,ReipurthMikkola2012,Reipurth2015}.     \citet{Goddi2011,MoeckelGoddi2012} assumed that Source I is a compact binary which along with BN were ejected from OMC1.  Their numerical simulations of three-body  interactions show that the most likely initial configuration is a pre-existing binary containing a massive star and BN, that experienced a close encounter with a massive star in which  the single star exchanged place with BN and ejected the latter.   The resulting binary must have a much smaller semi-major axis than the initial binary.      The recent measurement of the  proper-motion of Source n \citep{Rodriguez2016} suggests that a 4-th star was involved.   If source n is a wide binary as indicated by the radio data, then at least 5 stars were involved in the dynamical interaction.       

\subsection{The OMC1 Outflow}

OMC1 contains a wide opening-angle, arcminute-scale outflow traced by millimeter and sub-millimeter emission lines that exhibit  broad   ($>$ 100 km s$^{-1}$)  wings in molecules such as CO, SO,  and SiO \citep{Snell1984},  high-velocity  H$_2$O and SiO  maser emission \citep{Genzel1981,Greenhill1998,Matthews2010}, and hundreds of individual bow shocks in the near-IR lines of  \Htwo\  and  [\Feii ] dubbed the `Orion  fingers' \citep{AllenBurton93,Colgan2007,Youngblood2016}.    Over a dozen shocks protrude into the Orion Nebula as Herbig-Haro (HH) objects \citep{AxonTaylor1984,ODell1997a,ODell1997b,Doi2002,Kaifu2000,Doi2004,Graham2003,ReipurthBally2001} demonstrating that the outflow is located within $\sim$0.1 pc of the ionization front at the rear of the Nebula.    The  HH objects and [\Feii ] knots north and northwest of OMC1,  located at the largest projected distance from this core,  exhibit proper motions  of over 400 \kms ,  indicating a $\sim$500 year age of the outflow, similar  to the time since the ejection of the stars \citep{JonesWalker1985,LeeBurton2000,Doi2002,Bally2011,BallyGinsburg2015}. 

Observations of molecules show that the OMC1 outflow contains at least 8 M$_{\odot}$ of  accelerated gas in a roughly 0.2 pc ($6 \times 10^{17}$ cm) diameter region \citep{Snell1984}.  Half of the mass has an expansion velocity below 20 \kms\  with respect to OMC1;   the rest is traced by fading red- and blue-shifted line-wings extending to over 100 \kms\ with respect to the radial velocity of the emission from the Orion A cloud.     The outflow exhibits a slight elongation with red-shifted emission to the southeast and blue-shifted emission towards the northwest in the lower-velocity gas.    The Integral Shaped Filament that can be traced for nearly 1\arcdeg\  north-to-south behind the Orion Nebula \citep{JohnstoneBally1999} may partially block the outflow in the north-northeast and south-southwest direction. 

Interferometric  images of CO  emission with a resolution of about 3\arcsec\  obtained with the SMA  revealed several dozen, high-velocity streamers originating from OMC1 resembling a nearly isotropic explosion   \citep{Zapata2009c}.     These streamers exhibit  Hubble-law kinematics with radial velocities proportional to the projected distance from the location from which  BN and Source I were ejected $\sim$500 years ago, and point back to this location.   Comparison of the CO with near-IR images indicate that there is a close connection between the \Htwo\ shocks and CO filaments.   

A smaller, 8"  (3300 AU) long, collimated outflow  traced by CO, SiO, and H$_2$O  masers emerges along a northeast-southwest axis  from radio Source I,  orthogonal to the arc minute-scale, lower-velocity components of the main CO outflow.  This compact, and therefore young ($<$200 year old) outflow emerges nearly orthogonal to the proper motion of Source I  and a line connecting Source I to BN \citep{Plambeck2009,Zapata2012}.      The  present-day momentum and kinetic energy content of the OMC1 outflow complex is at least  $160$ M$_{\odot}$ km s$^{-1}$ and  $4 \times 10^{46}$  erg to  $4 \times 10^{47}$ erg \citep{Snell1984}.   

We present new ALMA observations with $\sim$1\arcsec\ angular resolution covering a 2\arcmin\ by 3\arcmin\ region containing the entire OMC1 outflow.  We discuss the physical properties of the ejected gas and review a scenario which links the origin of the outflow to the dynamic ejection of massive stars from OMC1.

\section{Observations}

The OMC1 outflow was mapped with the Atacama Large Millimeter Array (ALMA) in the 1.3 mm atmospheric window (ALMA Band 6)  using four 1.875 GHz-wide bands centered at 216, 228, 231, and 233 GHz.   The spectral windows utilized a channel-width of 0.488 MHz or an effective resolution (after Hanning smoothing) of about 1 MHz, providing velocity resolution of 1.3 \kms .  These bands contain the J=2-1 \Tco\ and \co\  lines,  the 217.1 GHz line of SiO, the 219.95 GHz line of SO, and a host of other lines.  Here, we only present the \Tco\  observations.    A future paper will present data on the other species present in the four 2 GHz-wide bands.   The 12-meter ALMA antennae have a 27"  full-width-half-maximum (FWHM) primary beam at this wavelength.   The 2\arcmin  $\times$3\arcmin\  field of view containing the outflow was observed using two rectangular grids.  An 80\arcsec $\times$180\arcsec\  rectangular grid used 108 antenna pointings with a grid-spacing of 12.9\arcsec\   covering the northwest part of the outflow.  A 70\arcsec\ $\times$ 70\arcsec\  grid using 39 pointings covered the southeast portion of the outflow.  The field was re-observed with the twelve-element ALMA Compact Array (ACA) of 7-meter diameter antennas and the four 12-meter diameter total-power antennae to provide smaller interferometer baselines and zero-spacing data.    

The raw data were reduced using the Common Astronomy Software Applications package, CASA \citep{casa}.  The ALMA QA2 reduction included heavy flagging to remove strong spectral lines from the continuum, as well as the standard flux, bandpass, and gain calibrations.  The data presented here used the combined 12- and 7-meter data.   The {\it uv}-plane visibilities from the two data sets were combined using the CASA  package, imaged, and CLEAN-ed using a Briggs robust weighting parameter of 0.5, resulting in a synthesized beam with dimensions of 1.\arcsec 36$\times$0.\arcsec 78.  At the 414 pc distance of Orion \citep{Menten2007},  the linear resolution is approximately 400 AU.   The rms noise varies across the field because of strong extended emission from OMC1. Typical noise values in clean regions of the continuum image are 0.5 mJy. The noise is 5 mJy  in a typical spectral line channel, although in channels with strong emission from the OMC1 outflow the r.m.s. can rise to 0.5 Jy.  At the 230 GHz frequency of the J = 2$-$1 CO line,  the antenna temperature scale is given by $T^*_A = 20.6$ Kelvin/Jy  in our synthesized beam.     Thus the brightness temperature sensitivity ranges from about 0.1 K per channel in emission-free regions to about 10 K per channel in the channels containing the bulk of the CO emission from Orion A (radial velocities from  \Vlsr\  $\approx$ 8 to 12 \kms  ) and at high velocities within tens of arc-seconds of OMC1 where the CO lines are bright. 

\section{Results}

The  ALMA observations reveal a cluster of over 100 high-velocity CO streamers  that trace back to a location between the BN object and Source I.    Figure~\ref{fig1}  shows the maximum flux  (brightness temperature) in each pixel from  \Vlsr\ = $-$100 to +120 \kms ; Figure ~\ref{fig2} shows a moment 0 image over the same radial velocity range.    Figure~\ref{fig3} shows a color version of the maximum temperature from \Vlsr\ = $-$100 to 0 \kms\ in blue and from   \Vlsr\ = 20 to 120 \kms\ in red.    The maximum brightness temperature better illustrates the relatively constant  peak surface brightness along the streamers. Furthermore, in this presentation the data is only impacted by the noise in a single channel containing the brightest emission; in an integrated image, the noise increases as the square root of the number of channels in the displayed velocity interval.    
 
The peak CO flux in the streamers is about 4.3  Jy/beam in both the redshifted and blueshifted ranges which corresponds to brightness temperature of $\sim$90 K .    The streamers dominate the outflow structure at large radial velocities,  are straight with lengths ranging from 5\arcsec\ to over 40\arcsec\  ($\sim$0.08 pc), and point back to J2000 = 05$^h$35$^m$14.34$^s$, $-$5$^o$22'28.4", within a few arc-seconds of the location from which radio Source I and BN were ejected, within 1\arcsec\ of the CO ejection center determined by \citet{Zapata2009c}, and within a few arcseconds of the ejection center determined from a trace-back of the \Htwo\ proper motions \citep{Bally2011}.   Most streamers are barely resolved with widths ranging from 1\arcsec\  to 2\arcsec .   They exhibit low-amplitude intensity variations indicating clumpy structure and arcsecond-scale wiggles.   The line-widths of the streamers range from slightly under 5 \kms\ to slightly over 10 \kms\ (full-width-half-maximum; FWHM) in a synthesized beam area.         The brightest streamers, those with T$_B >$ 10 K  are confined to a $<$50\arcsec\ radius circle centered at J2000 = 05:35:14.1, $-$5:22:19, about 10\arcsec\ north-northwest of the ejection center (Figure \ref{fig1b}).    Fading emission with T$_B <$ 10 K can be traced farther north towards some of the \Htwo\ fingers and towards the southeast. 

\subsection{High radial-velocity structure}

\begin{figure}
\begin{center}
\includegraphics[width=6in]{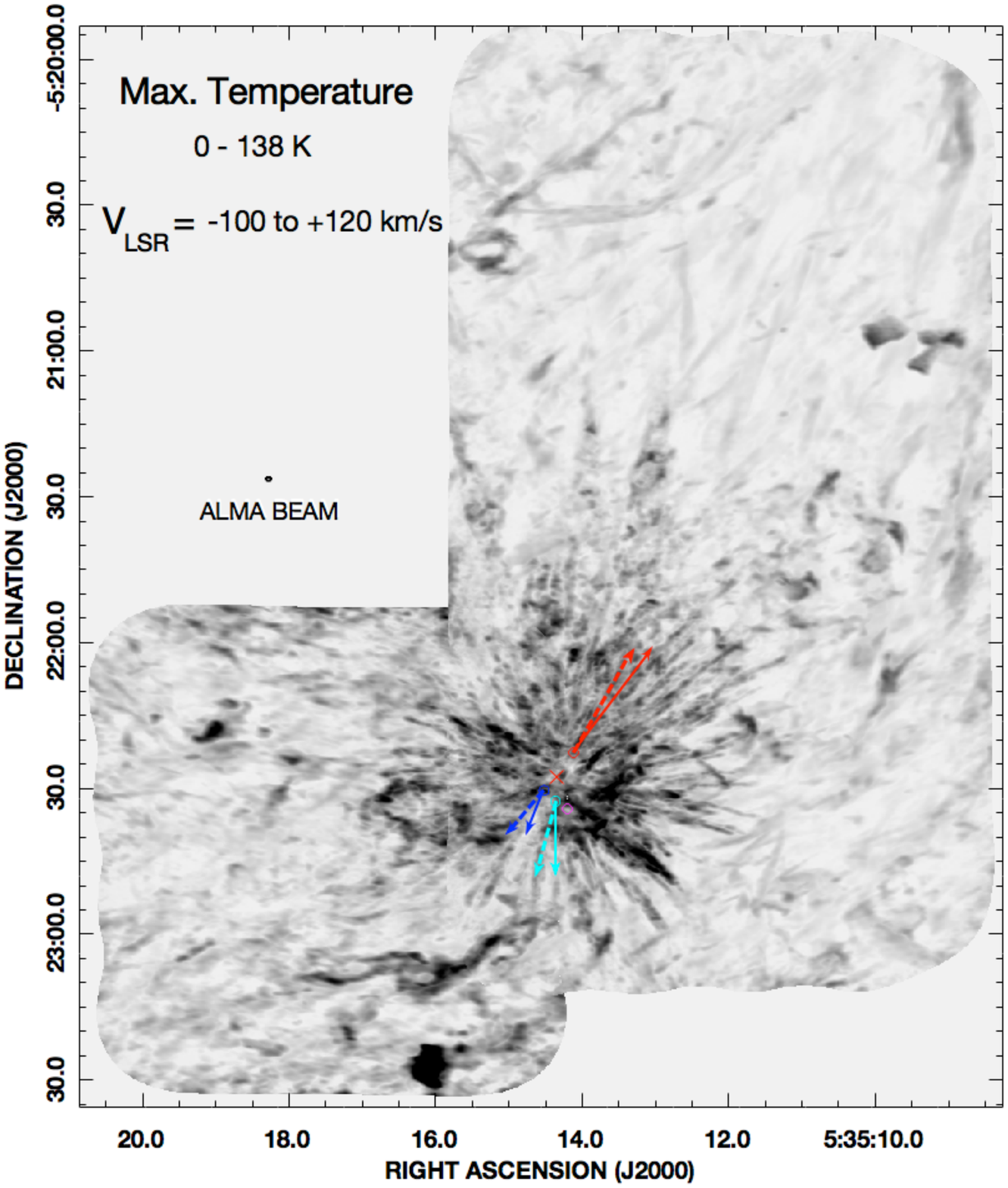}
\end{center}
\caption{An ALMA 1.3 mm image showing the high-velocity 230 GHz CO emission from the explosive OMC1 outflow with $\sim$1" ($\sim$400 AU) angular resolution.  The maximum brightness temperatures  in velocity channels covering the indicated velocity intervals are shown from $\rm T_A^*$ = 0 to 89 K on a linear scale.  The current positions of BN, Source I, Source n, and IRc4 are indicated by red, blue, cyan, and magenta circles respectively.  The vectors show the proper motions of BN (red), Source I (blue), and Source n (cyan)  in the Orion Nebula Cluster reference frame \citet{Dzib2016,Rodriguez2016}.   The dashed vectors show these motions in a frame in which the vector momenta sum to zero as described in the text.  The lengths of the vectors correspond to the expected proper motions in 2,000 years.  
}
\label{fig1}
\end{figure}

\begin{figure}
\begin{center}
\includegraphics[width=6in]{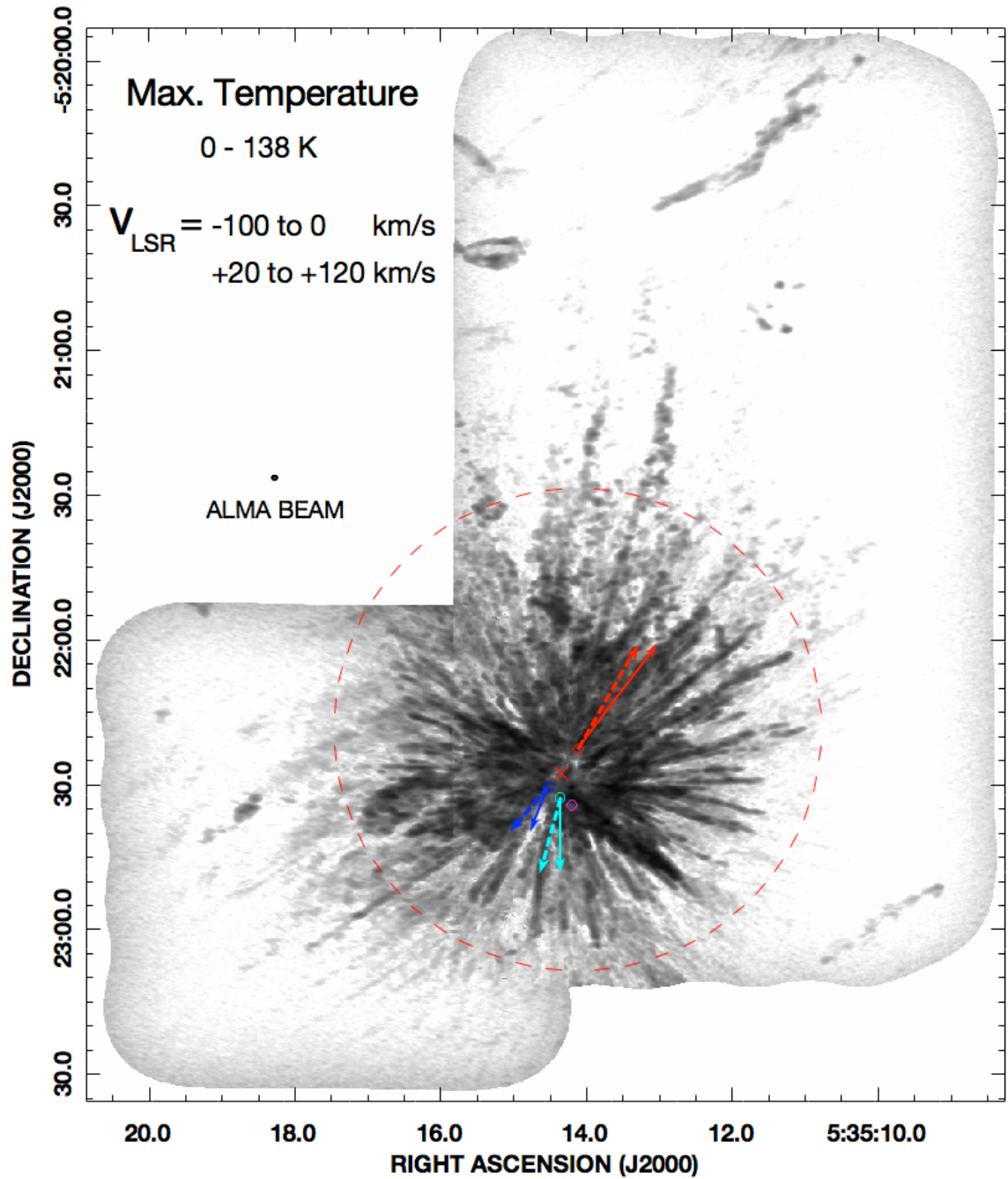}
\end{center}
\caption{An ALMA CO image showing only the high-velocity emission, excluding the velocity range from\Vlsr\ = 0 to +20 \kms .  As in Figure \ref{fig1}, the maximum brightness temperature in the indicated velocity ranges is shown.    However, in this figure, the logarithm of $T_{max}$ is shown to emphasize the low-level emission.   A dashed circle with a radius of 50\arcsec\ enclosing most of the high-velocity CO emission is shown for reference.  The center of this circle is at J2000 = 05:35:14.11, -5: 22:18.7, about 10\arcsec\ north-northwest of the ejection center.   Proper motion vectors and luminous sources are shown as in Figure \ref{fig1}.   Note the small outflows in the lower-right and top of the figure.  These are discussed in Section 3.6.
}
\label{fig1b}
\end{figure}

\begin{figure}
\begin{center}
\includegraphics[width=6in]{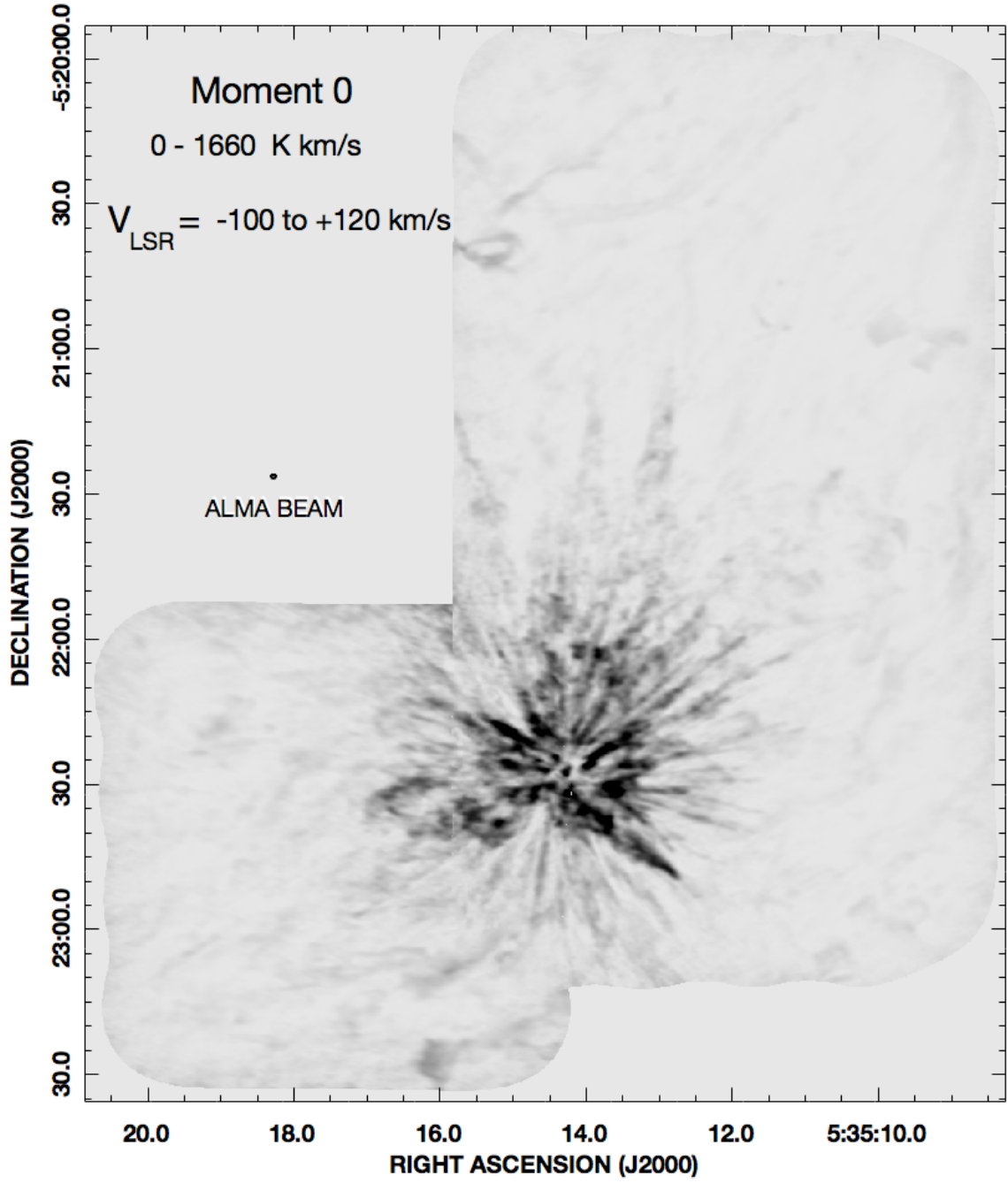}
\end{center}
\caption{A moment 0 map showing the high-velocity 230 GHz CO emission from OMC.  }
\label{fig2}
\end{figure}

\begin{figure}
\begin{center}
\includegraphics[width=6in]{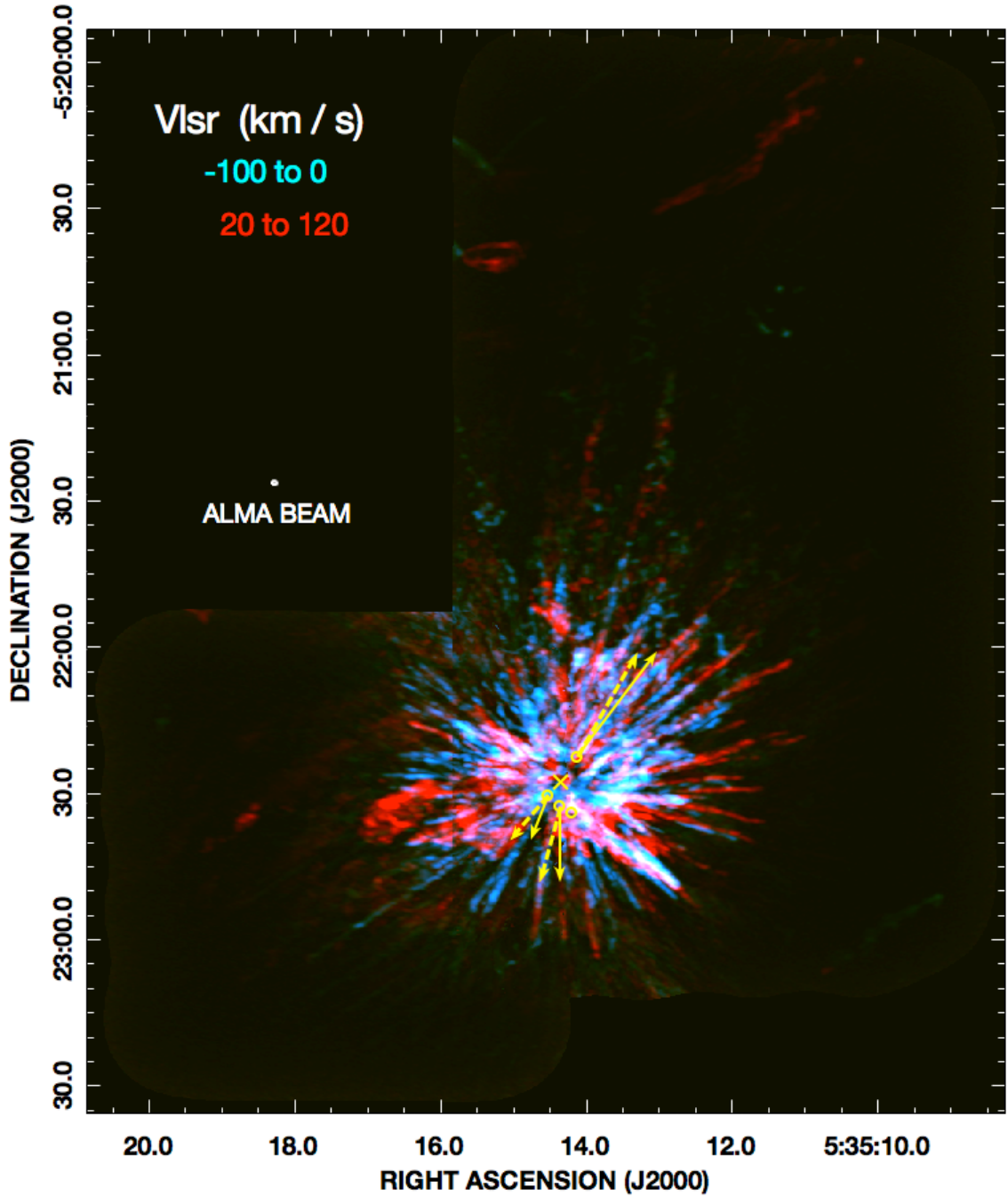}
\end{center}
\caption{The blue-green features show the outflow emission  from \Vlsr  =$-$100 to 0 \kms ;  red features show  emission from  \Vlsr  =+20 to +120 \kms .   The yellow circles show the locations of the massive stars ejected from OMC1;  from top to bottom these are the BN object, Source I, Source n, and IRc4.   The yellow cross  marks the  location of the explosion center measured from the proper motions of near-infrared \Htwo\  emission features \citep{Bally2011};  this location coincides with the explosion center determined from the orientations of the CO streamers to within a few arc-seconds \citep{Zapata2009c}.   Several unrelated CO outflows from stars outside the OMC1 core can be seen near the top and lower-right corner of this image.
}
\label{fig3}
\end{figure}

At radial velocities greater than 20 \kms\ with respect to the \Vlsr\ = 9 \kms\ velocity of OMC1, over 95\% of the high-velocity CO emission originates from about 100 streamers confined to within $\sim$0.1 pc of the ejection center.  This region coincides with the brightest near-IR \Htwo\ emission from Peak 1 and Peak 2 \citep{Youngblood2016}.   As shown in Figures~\ref{fig1},  \ref{fig1b}, \ref{fig2}, and \ref{fig3},  the streamers have an approximately spherically symmetric spatial distribution.   However, there are some significant deviations from this symmetry as discussed below.

\begin{figure}
\begin{center}
\includegraphics[width=3.5in,angle=-90]{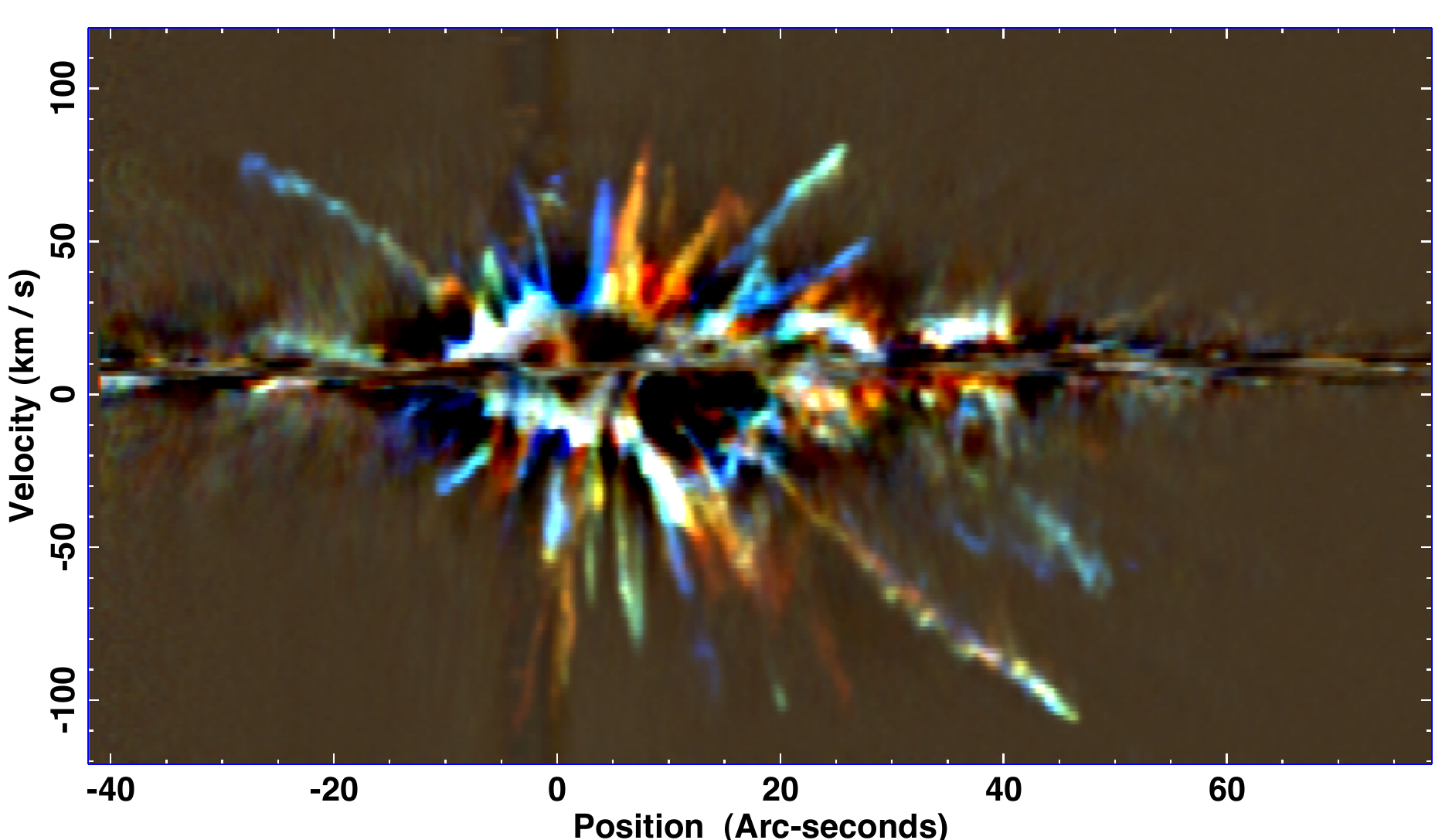}
\end{center}
\caption{ North-south cuts across the OMC1 outflow at the Right Ascension of the explosion center showing the radial velocity structure  over a 80" ($\sim 5 \times 10^{17}$ cm or 0.16 parsec) spatial region centered on the outflow in the 230 GHz CO J=2$-$1 emission line.  Three north-south strips (equivalent to `long-slit' spectra) separated by 0.6" are shown in red, green and blue.  The green channel is centered at R.A.= 05$^h$35$^m$14.$^s$364.   The red channel shows the `long-slit' spectrum 0."6 to the east of this position; the blue channel shows the `long-slit' spectrum 0."6 west of this position.  The intensity scale ranges from $-$10 to 80 K. {\bf An animated version of Figure \ref{fig4} showing the long-slit spectrum (upper panel) in which the slit orientation pivots 360 degrees around the explosion center (lower panel) is provided in the online Journal. The animation is 15 seconds in duration.}
}
\label{fig4}
\end{figure}

Figure~\ref{fig4}  shows a north-south cut through the ALMA CO data cube illustrating the velocity behavior of several dozen north-south oriented streamers.    The streamers exhibit large-amplitude, linear-velocity gradients, a result previously demonstrated by \citet{Zapata2009c,Zapata2011a,Zapata2011b}.  An animated version of this figure illustrating the spatial-velocity behavior of the streamers at all position angles with respect to the ejection center is shown in the electronic version of this article.  The streamers are mostly straight in both the images and in spatial-velocity cuts through the data cube;  each consists of a stream of clumps moving along a well-defined direction with radial velocities increasing in proportion to the projected distance from the ejection center (e.g. a `Hubble flow').    The radial velocity structure of the streamers at all orientations from the dynamical ejection center shows similar behavior.  The typical radial-velocity gradients range from less than 200 to over 2,000 \kms\ per parsec.     In some, such as the 40\arcsec -long streamer shown in lower-right of Figure~\ref{fig4}, the line-brightness peaks near the high-velocity end, implying either enhanced heating, or a concentration of mass at the leading, high-velocity edge.     Some streamers appear brighter on their high-velocity sides.   The streamers exhibit little deceleration, implying that their densities are much higher than the ambient medium through which they are moving.     However, some streamers contain faint wisps of CO emission extending vertically in Figure~\ref{fig4} to lower radial velocities.   They may trace gas being shorn off and decelerated as the dense clumps move supersonically through the medium.     

Most streamers intersect the 6 to 12 \kms\ rest velocity (in \Vlsr\ coordinates) of the Orion A cloud within a few arc-seconds of the ejection site of Source I and BN.  However,  extrapolating the linear velocity-gradient of the two longest streamers  in Figure~\ref{fig4} (one in the upper-left of Figure~\ref{fig4} and the other being the 40\arcsec -long streamer in the lower-right) back to the velocity of the OMC1 core and the Orion A cloud results in intersection points separated by $\sim$5 \arcsec .     Fitting lines to the streamers in the spatial-velocity images also results in a spread of intersection point scattered over a roughly 5\arcsec\ diameter region.  This behavior may indicate slight deflections by interactions with their environment or spreading of the ejecta .  Nevertheless the distribution of intersection points in the space-space images (Figures \ref{fig1} to \ref{fig4}) peak within an arcsecond of the dynamical ejection center determined from proper motions of the stars referenced to the frame in which OMC1 is at rest, as discussed below. 

Dozens of red- and blueshifted CO streamers point towards the \Htwo\ and [\Feii ] bow shocks.  This is particularly evident towards the southwest.   Although some CO streamers extend right up to associated \Htwo\ bow shocks located at their high-velocity ends,  others  disappear between 30 to 70\% of the distance to the near-IR bow shocks to which they point.   Given that the HH-object and some near-IR proper motions reach values of $\sim$400 \kms , a factor of two to three times the highest CO radial velocities, it is likely that CO and \Htwo\ are dissociated by the intense shock heating and radiation above speeds of order 100 to 150 km/s.  

The dynamical ages of the streamers can be estimated from their maximum radial velocities if their inclination angle with respect to the line-of-sight,  or their proper motions are known.    Assuming spherical symmetry and that all streamers were produced at the same time, the maximum projected distance of the streamer-ends from the ejection center can be combined with their radial velocities  to estimate their inclination angles and ages.   Most bright streamers are confined to within $R_{max}$= 50\arcsec\ ($3.1 \times 10^{17}$ cm) of the explosion center.    The highest radial velocities are $V_{max}$ = 159 \kms\  with respect to OMC1.  (Knot \#1 in Table~1 has \Vlsr\ = $-$149.5 \kms .  Since OMC1 is at \Vlsr\ = +9 \kms , this knot is moving with a velocity of at least 159 \kms\ with respect to the cloud core).   If the proper motions at 50\arcsec\ projected distance were 159 \kms , the dynamic age would be $t_{dyn} = R_{max} / V_{max} \approx$ 620 years.
     
Inclination angles  of individual streamers can be estimated from their maximum radial velocity, $V_{radial}$, and projected distance of their high-velocity end,  $r_{end}$ from the ejection  center.   The ratio $r_{end} /  R_{max}$ is a measure of the inclination angle of the streamer with respect to the line-of-sight.    Thus,  
$ 
 t_{dyn} \approx  ~ (r_{end} / V_{max})~ tan~[~cos^{-1}(r_{end} / R_{max})].
$ 
For example, the long streamer in the lower-right of Figure~\ref{fig4} has a projected length of $r_{end} \approx $ 40\arcsec\ and a radial velocity with respect to OMC1, $V_{radial} \approx $110 \kms , at its high-velocity end.  For R$_{max}$=50\arcsec , its dynamical age is $t_{dyn} \approx $540 years.  The ALMA data are consistent with an age of 470 to 700 years,  similar to the age obtained by \citet{Zapata2009c} with the SMA and the recent analysis of the time since ejection of the high-proper motion stars from OMC1 by \citet{Dzib2016,Rodriguez2016}.    The fastest blueshifted and redshifted  knots are displaced from the ejection center in projection, indicating some deviations from spherical symmetry.  Furthermore, as discussed below, many of the \Htwo\, [\Feii ], and HH objects in the north and  northwest are located up to 140\arcsec\ from the ejection center.  

\subsection{Low radial-velocity structure}

Within $\sim$20\arcsec\ of the ejection site, the low-radial velocity CO emission within $\sim$20 \kms\ of the OMC1 rest-velocity (the emission line from the core is centered at \Vlsr [core]\ $\approx$ 9 to 10 \kms ) becomes over-resolved in the ALMA data.   The outer boundary of this region is seen as an approximately ovoidal shell with a chaotic boundary (Figure \ref{fig4}) with an interior region with negative intensities.    The combined 12-meter and ACA data over-resolves this region and thus there is a large amount of missing flux.    Nevertheless  an abundance of  small-scale structure can be seen at low radial velocities and beyond 20\arcsec\ of the core.

\begin{figure}
\begin{center}
\includegraphics[width=6in]{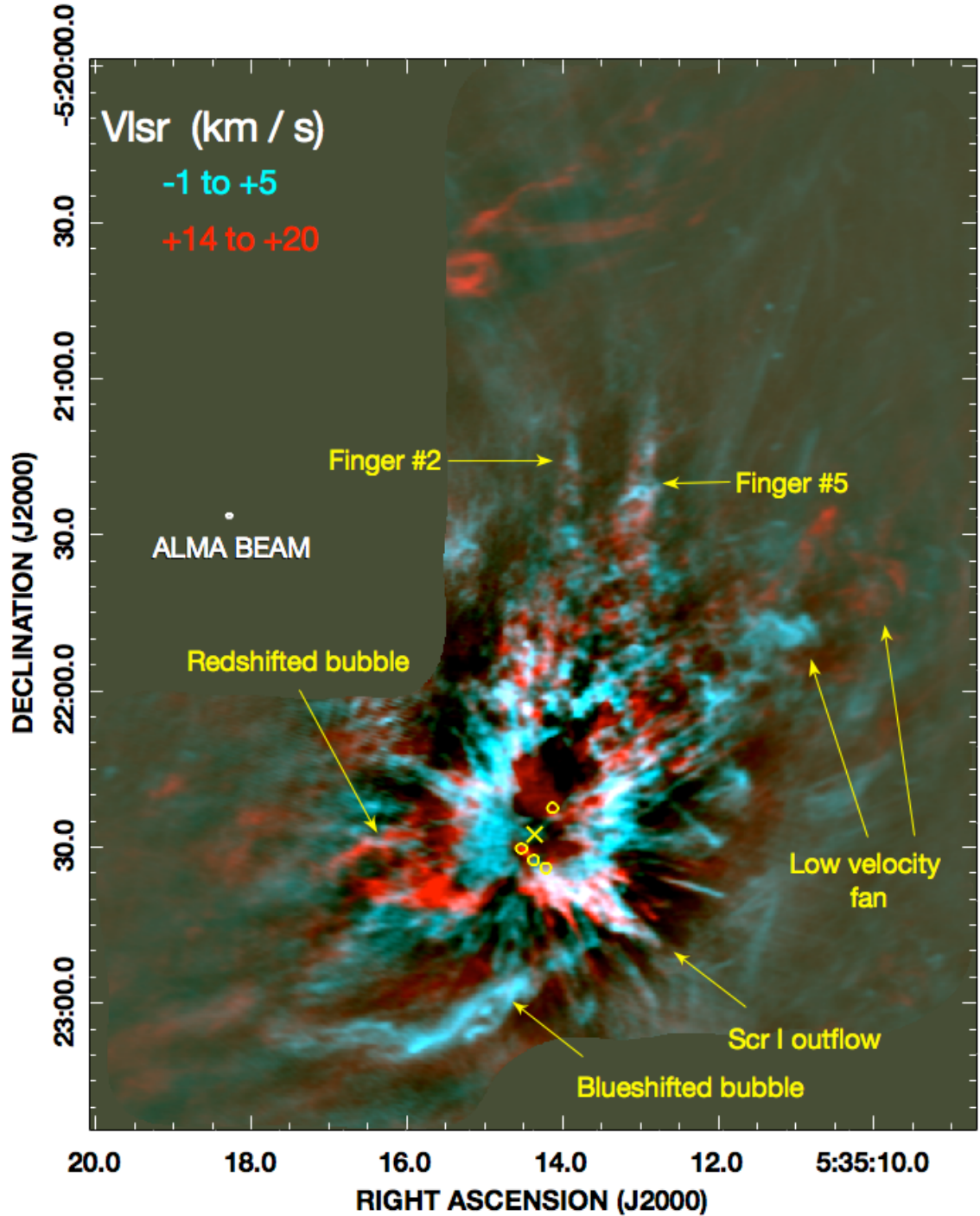}
\end{center}
\caption{An ALMA 1.3 mm image showing the  230 GHz CO J=2$-$1 emission at low radial-velocities formed by integrating the line intensity over a velocity range   from \Vlsr  =$-$1 to +5 \kms\  (shown in cyan)  and  \Vlsr  = +14 to +20 \kms\  (shown in red).    The ejected massive stars and outflow ejection center are shown as in Figure \ref{fig3}.   Several additional outflows from young stars in the Orion A cloud can also be seen in this velocity range towards the north side of the image.   The displayed intensities range from -7 to 63 K.
}
\label{fig5}
\end{figure}

Within $\sim$10 \kms\ of the rest velocity of OMC1, the outflow is elongated from southeast to northwest with a length of nearly 180\arcsec\  (0.36 pc: Figure~\ref{fig5}) and contains several shell-like structures.      An arcminute long, 30\arcsec\ wide, blueshifted bubble with walls several arcseconds-wide ($\sim$1,000 AU) extends southeast at low blueshifted radial velocities of \Vlsr $\sim$ 0 to 7 \kms ; its  southern rim  is located behind the Trapezium stars.   A slightly smaller, redshifted bubble protrudes due east of the OMC1 cloud core at \Vlsr\ $\sim$12 to $\sim$55 \kms .      A broad fan of low-velocity red and blueshifted ejecta extends about 1.5\arcmin\ towards the northwest, in a direction opposite to the southeast oriented blueshifted bubble.   These bubbles are marked in Figure~\ref{fig5}. 

\begin{figure}
\begin{center}
\includegraphics[width=6in]{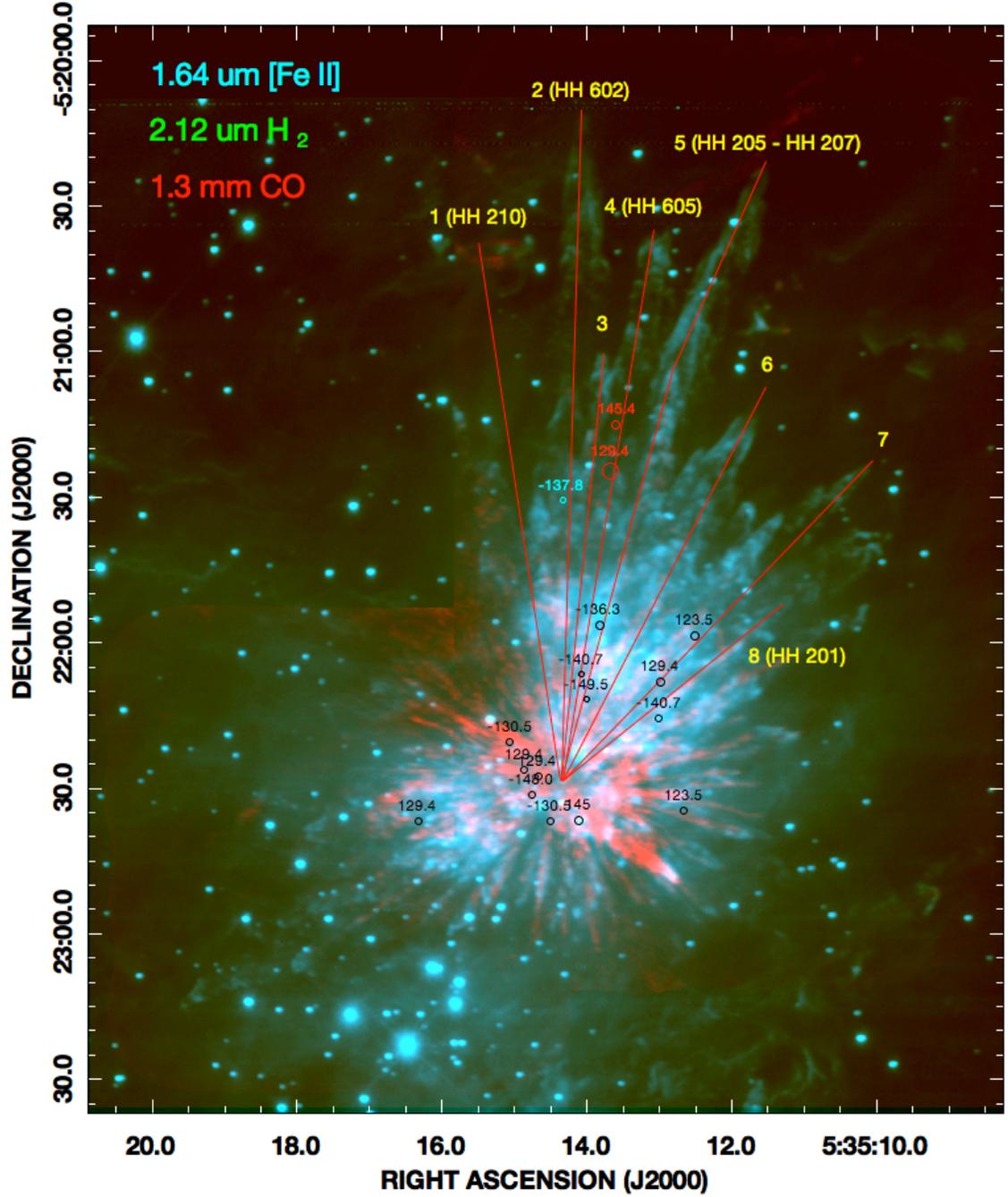}
\end{center}
\caption{The maximum brightness temperature CO maps in the redshifted  and blueshifted radial velocity ranges shown in Figure 1 covering the high-velocity emission from $-$100 to +120 \kms\ (red)  shown superimposed on the 2.12 $\mu$m fingers of shock-excited \Htwo\ emission (green)  from the OMC1 outflow \citep{Bally2011}.  The massive Trapezium stars responsible for ionization of the Orion Nebula appear near the bottom of this image.    The eight most prominent \Htwo\ fingers are marked with red lines.  The HH objects located at their ends are indicated.   The small red and black circles indicate the locations and radial velocities of the highest velocity streamer ends listed in Table~1.
}
\label{fig6}
\end{figure}

Several limb-brightened `fingers'  of CO emission extend north and northwest for 60\arcsec\  to 130\arcsec\  at low-radial velocities.  CO emission is clearly associated with the \Htwo\ fingers  \#2, 3, 5, and 8 (Figure~\ref{fig6}).    Both redshifted and blueshifted CO with velocities of $\pm$3 to 20 \kms\ with respect to \Vlsr =  9 \kms\ are associated with the \Htwo\ fingers  \#2 and \#5, suggesting that they lie close to the plane of the sky.   Low-radial velocity material in fingers \#3 and \#5 is located in the interior of the \Htwo\ wakes, and likely traces molecules moving at high velocities nearly in the plane of the sky (Figure~\ref{fig6}).     The elongation of the low-velocity bubbles and the fingers are roughly  parallel to the motions of BN and Source I.   

\begin{figure}
\begin{center}
\includegraphics[width=6in]{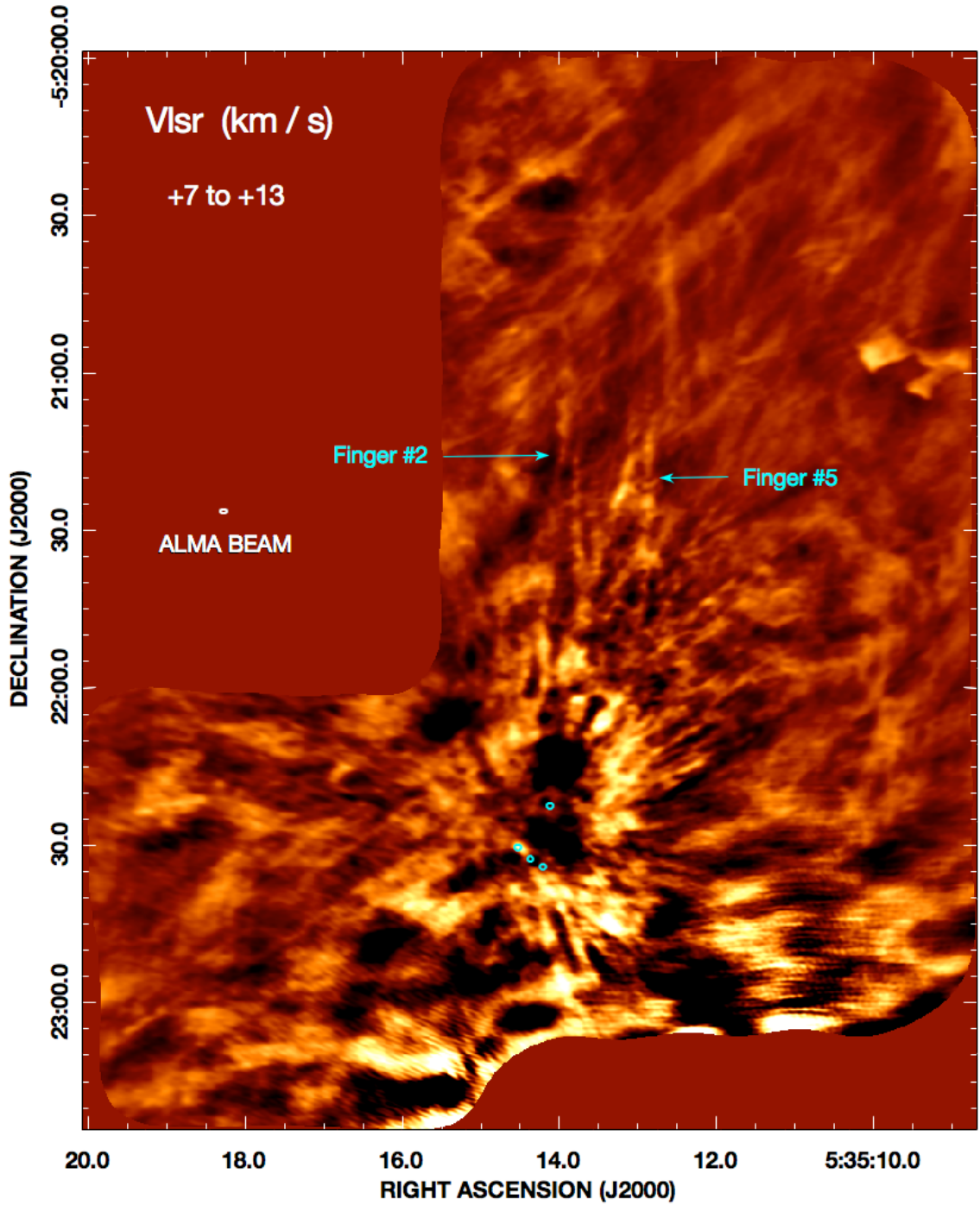}
\end{center}
\caption{An integrated CO 2$-$1 image illustrating the small angular-scale structure from \Vlsr = 6 to 13 \kms .   Comparison with Figures \ref{fig3} and \ref{fig4} shows that at these velocities, cones of CO emission originate from outside the prominent \Htwo\ streamers 3 and 5.   The patchy dark bays surrounding the ejection center (plus-symbol) and other parts of the image indicate missing flux due to spatial filtering of the ALMA data.    Blue circles (from north to south) mark the locations of BN, Source I, Source n, and IRc4. 
}
\label{fig7}
\end{figure}

Figure~\ref{fig7} shows the integrated  CO emission at the core radial velocities, \Vlsr = 6.7 to 12.6 \kms .  Conical ridges of enhanced CO emission best seen in Figure~\ref{fig7} are located just outside  \Htwo\ fingers 3 and 5 and may trace limb-brightened ambient gas that has been displaced and slightly accelerated to redshifted velocities by the passage of high-velocity ejecta.    At the core velocities, the CO emission from  fingers \#2 and \#5 consists of a pair of converging filaments located just outside the associated near-IR \Htwo\ emission.   These `upside-down V-shaped' features may trace very low-velocity CO swept-up from the ambient medium.   

\begin{figure}
\begin{center}
\includegraphics[width=6in]{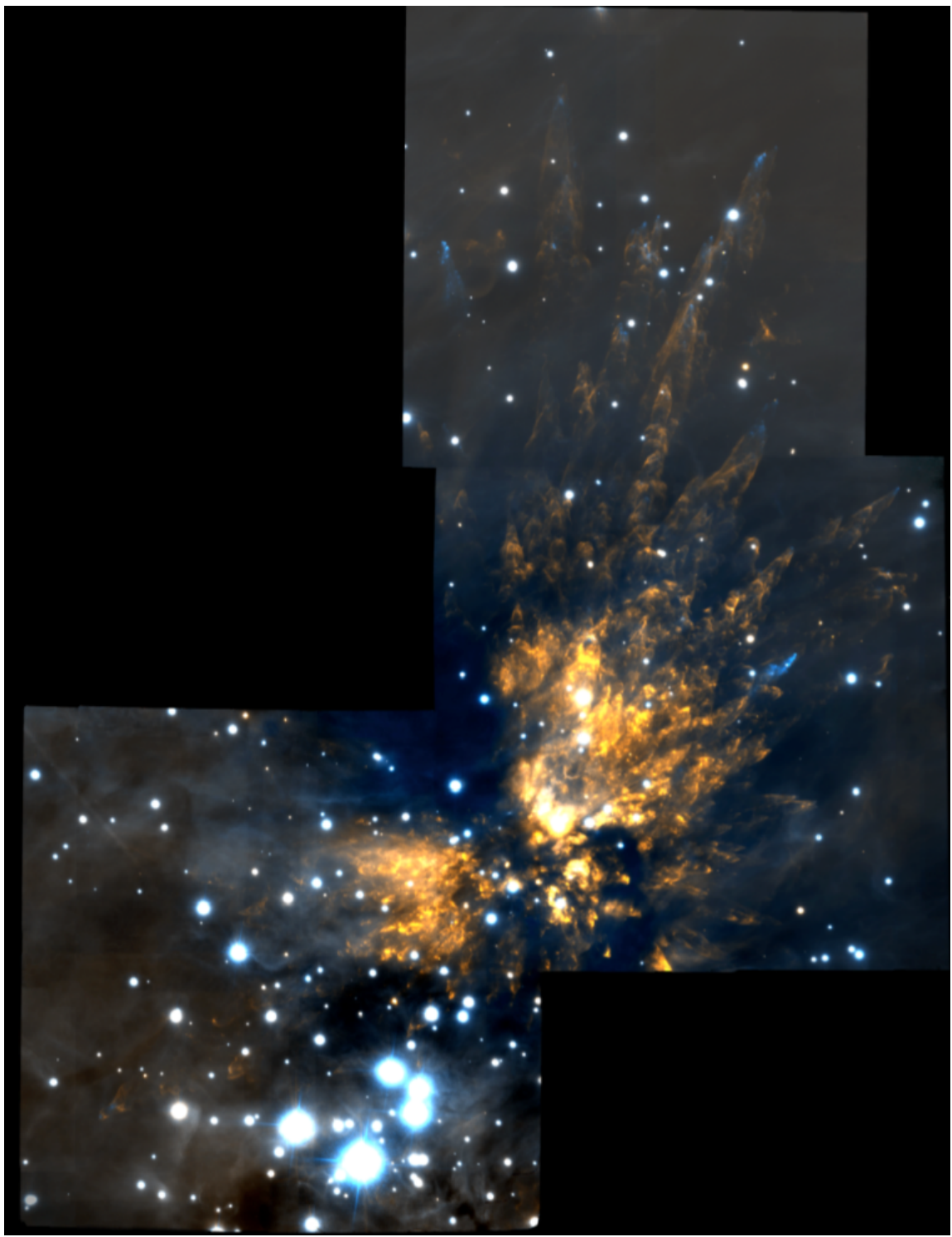}
\end{center}
\caption{The OMC1  outflow in the 2.12 $\mu$m \Htwo\ (orange) and 1.64 $\mu$m [\Feii ] (blue) lines in the near-infrared.  This image was obtained with the Gemini South Adaptive Optics Imager (GSAOI) and the Gemini Multi-conjugate adaptive optics System (GeMS) on the Gemini South 8-meter telescope that uses five Sodium lasers to generate artificial guide stars.   The angular resolution of these observations is about 0.06".  Taken from \citet{BallyGinsburg2015}. 
}
\label{fig8}
\end{figure}

\begin{figure}
\begin{center}
\includegraphics[width=8in]{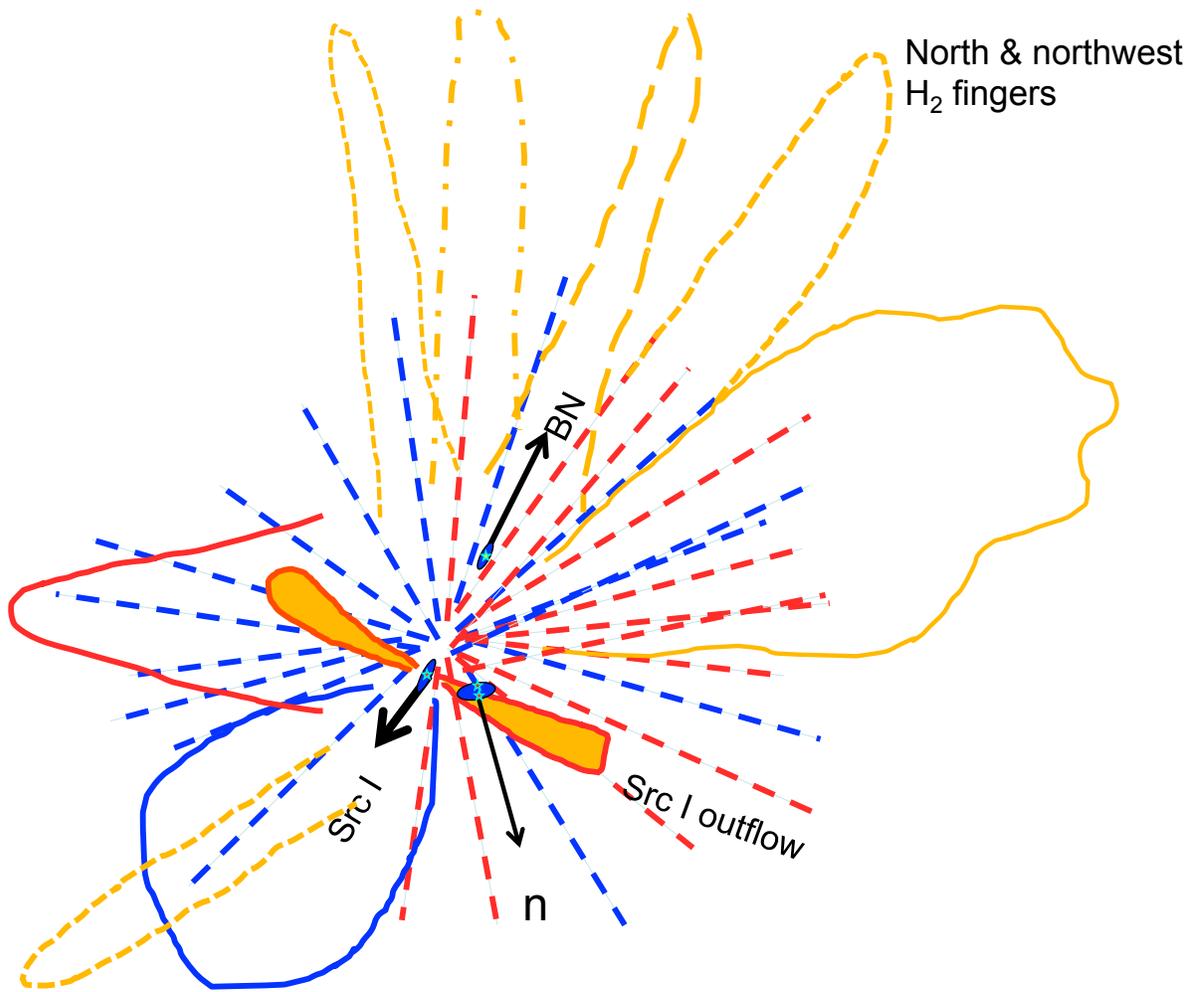}
\end{center}
\caption{A cartoon showing the major elements of the OMC1 outflow.  Dashed red and blue lines illustrate the high-velocity streamers.  Dashed yellow-orange lines mark the locations of the most prominent \Htwo\ fingers;  the solid orange line shows the bubble of low-velocity red-and blueshifted ejecta fan toward the northwest.  Solid red and blue lines show the red and blueshifted bubbles of ejecta extending to the east and southeast.  The solid orange region bounded by a solid red line shows the young Source I flow.  The relative locations of Source I, n, and BN, their proper motions, and suspected orientations of their disks are also illustrated, but not to scale.  
}
\label{fig9}
\end{figure}

Figure \ref{fig8} shows a high resolution image of the outflow from \citet{BallyGinsburg2015}.  Figure  \ref{fig9} shows a cartoon illustrating the various features discussed above.

\subsection{Deviations from spherical-symmetry}

The redshifted  emission is dominated by the low-radial-velocity bubble pointing due-east of the OMC1 core and a spray of several dozen high-velocity streamers oriented toward the west.   The reshifted bubble has an axis of symmetry pointing towards PA $\approx$ 95\arcdeg\ from OMC1.  Except for the young and bright Source I outflow and a single redshifted streamer at PA $\approx$ 70\arcdeg , there are no redshifted  streamers between PA = 25\arcdeg\ and  170\arcdeg .   In contrast, the blueshifted streamers exhibit a more isotropic distribution.  

The largest observed radial velocities in CO are at \Vlsr\ = $-$142 \kms\   58\arcsec\ north of the ejection center, and +133 \kms\ about 74\arcsec\  north, significantly outside the 50\arcsec\ radius region containing most of the bright, high-velocity CO streamers.  Table~1 lists some of the highest velocity streamer ends;  their locations are marked in Figure~\ref{fig6}.   The most prominent \Htwo\ and [\Feii ] fingers are marked with red lines in Figure~\ref{fig6} where they are numbered 1 through 8.   The proper-motions of several of these fingertips reach V $\sim$400 km/s in \Htwo\ and [\Feii ], nearly three times faster than the highest radial-velocity CO streamers.  

The low radial-velocity CO emission extends outside a 50\arcsec\  radius circle containing most streamers.    Additionally, streams of ejecta extend to the north and northwest where the highest proper motion near-infrared `fingers' of \Htwo\ and `fingertips' of [\Feii ] are found.    As discussed below, both red- and blue-shifted low-velocity CO is associated with the bases of several of these fingers (Figure \ref{fig5}), indicating that there is an unusually fast fan of ejecta moving close to the plane of the sky towards the north and northwest.    

Models of binary in-spiral indicate ejection of debris along the orbital plane \citep{SokerTylenda2006}.  Given the proper motions of the BN object and Source I, the most likely final encounter direction is along a southeast-northwest axis.  Ejecta  launched during the in-spiral phase are expected to be launched preferentially in  this direction.   The observed low radial-velocity outflow and the fan of high-proper-motion ejecta is  elongated along the axis defined by the stellar proper motion vectors,  consistent with the ejection of material in the encounter plane.  These asymmetric features are marked in the cartoon shown in Figure \ref{fig9}.

As discussed above, a 30\arcsec\ wide by 70\arcsec\ long bubble of limb-brightened, blueshifted emission extends towards the southeast with an axis of symmetry at PA $\sim$ 130 to 140\arcdeg .     Figure \ref{fig5} shows that at low radial velocities, there is a wide-angle fan of both red- and blueshifed debris extending to about 90\arcsec\ from OMC1 at PA $\sim$310\arcdeg , roughly opposite to the blueshifted bubble.

The brightest  CO emission away from the OMC1 rest velocity is associated with the compact, $\sim$200 to 300 year old, northeast-southwest ejecta from source I \citep{Zapata2012}.   Although this component is difficult to separate from the other streamers in the 1.3 mm CO data cubes, it is very apparent in a number of other molecules such as SiO and even \co .  Thus, Source I continues to power a very dense bipolar outflow roughly orthogonal to its proper motion.

\subsection{Column densities and masses}

Far-infrared observations of high rotational states of CO with the Herschel space telescope \citep{Goicoechea2015} show that the spectral line energy distribution (SLED) can be well fit with three components having temperatures of 200 K, 500 K and 2500 K.  The ALMA data cube shows that most streamers have brightness temperatures between 20 to 90 K and Doppler widths of $\Delta V \sim$ 5 to 10 \kms.    Assuming cylindrical geometry,  a characteristic width or depth of 1.2\arcsec\ (500 AU), and a \Tco\ abundance $X_{CO} = 10^{-4} n(H_2)$, the relation between column density of CO and the density of \Htwo\ is $N_{CO} = 7.5 \times 10^{11} n(H_2)$ cm$^{-2}$.   

The column density and density of the streamers depends on the assumed gas and radiation temperatures.  If the CO emission we observe originates in the 200 K component and the background radiation temperature is 10 K,  appropriate for the outer parts of the OMC1 outflow, the radiative transfer and excitation code RADEX \citep{vanderTak2007} indicates that the CO excitation temperature in the J=2$-$1 line ranges from 25 K to 88 K for \Htwo\ densities $n(H_2) = 4 \times 10^5$ to $2 \times 10^6$~cm$^{-3}$ for $\Delta V \sim$ 5 \kms\ wide lines.    For a background radiation temperature of 100 K, appropriate for regions close to (within a few arcseconds) the massive stars, the observed CO brightness requires densities of $10^6$ to $10^7$~cm$^{-3}$.  The CO opacities range from 0.3 to 3 for both cases.   These optical depths are consistent with the non-detection of \co\ at a brightness temperature above 0.2~K (10 mJy/beam) at high radial velocities.   Given the peak brightness temperature in the streamers of $\sim$90~K, the lowest allowed kinetic temperature is 100 K.  In this case, the acceptable volume density ranges from $\sim 10^6$ to $\sim 10^7$~cm$^{-3}$ for a background radiation temperature of 10~K.  For line-widths of 10 \kms , the line opacities are approximately a factor of two lower and the densities required to produce the observed line brightness are about a factor of two higher than for the estimates for 5 \kms\ wide lines.  It is likely that there is a wide range of gas kinetic temperatures in each streamer with hot $\sim 10^3$ K gas near their leading edges and sides where shock radiation and compression heat the post-shock gas to below 200 K in their interiors where radiative cooling dominates.    Lowering the CO abundance increases the required density.    

Column densities can also be estimated using the X-factor method \citep{Bolatto2013} which gives  $N(H_2) \approx ~2.6 \times 10^{20} ~T^*_A \Delta V$ where $T^*_A$ is the antenna temperature corrected for atmospheric attenuation and $\Delta V$ is the effective width of the CO spectra.    The conversion between the CO flux in Jy per beam and mass of hydrogen in the beam is $M(H_2) \approx 1.2 \times 10^{30} ~S_{beam}(Jy)$ grams ($\sim 6 \times 10^{-4} ~S_{beam}(Jy)$ \Msol ) where $S_{beam}(Jy)$ is flux density in a 1.46 \kms -wide channel.   The mass contained in each $\sim$1" resolution element ranges from $\sim 10^{29}$ to slightly over $3 \times 10^{30}$ grams.  Summing over the spatial and velocity extents of the filaments gives masses ranging from a few times $10^{29}$  grams for faint features only a few arc-seconds long to over $10^{32}$ grams for the longest streamers.   The total mass of the streamers traced by CO having a radial velocity that differs from the ambient cloud by more than 20 \kms\ is $M_{fast} \sim $1 \Msol . 

Integration of the CO data cube over an 86" diameter circle and summing the flux over the radial velocity ranges $-$100 to 0 \kms\ and +20 to 120 \kms\  implies a mass of 1.0 \Msol\ using this method.  The spatial-filtering of the interferometer severely limits our ability to estimate masses within the inner $\pm$10 \kms\ of the outflow;   even at higher velocities, extended emission is filtered out and thus our mass estimates are lower bounds.    

\subsection{The relationship of the CO streamers and  the \Htwo\ fingers}

The \Htwo\ fingers and [\Feii ] bow shocks extend $\sim$140\arcsec\ from the ejection center towards the north and northwest (Figure \ref{fig8}).    The multi-transition analysis of the \Htwo\ in the  OMC1 outflow \citep{Youngblood2016} shows that near the core, the  \Htwo\ emission is primarily blueshifted while towards the outer parts towards the north, northwest, and the southeast, the \Htwo\ emission is both red- and blueshifted.   \citet{Oh2016} present high spectral resolution data cubes of \Htwo\  towards a 15\arcsec\ by 13\arcsec\ field centered on the brightest portion of the Orion outflow $\sim$25\arcsec\ north-northwest of the ejection center.  The \Htwo\ emission in this field has morphology and kinematics similar to the CO streamers in the region with red and blueshifted emission extending from $-$105 to +95 \kms .

\begin{figure}
\begin{center}
\includegraphics[width=6in]{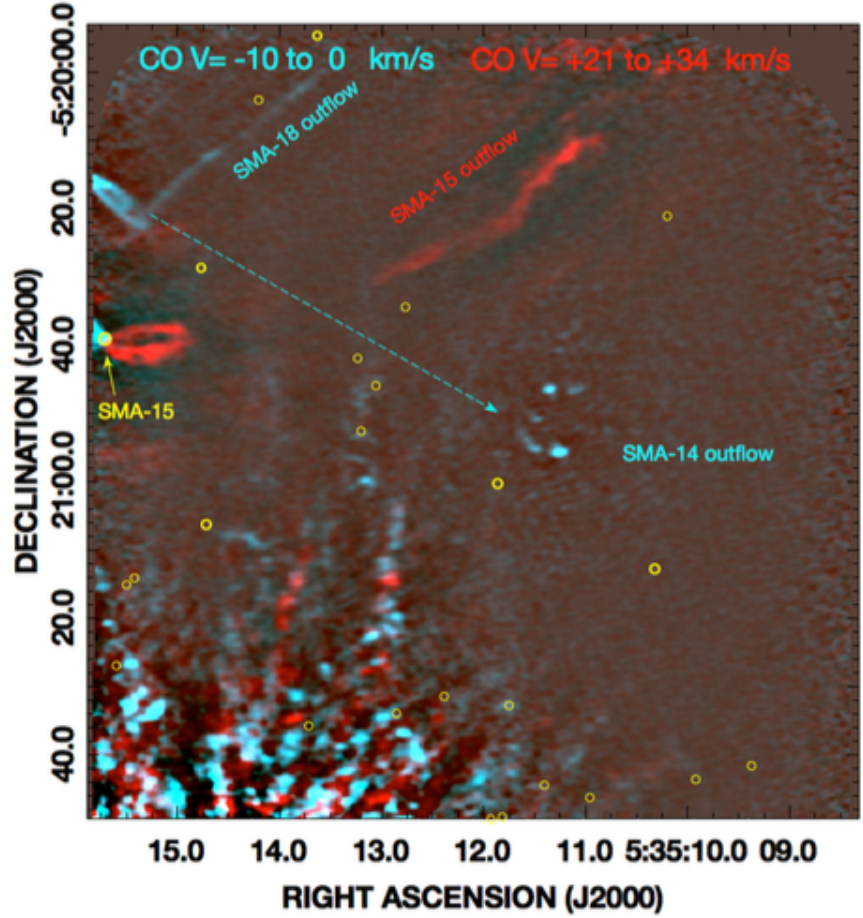}
\end{center}
\caption{An image showing the CO emission at \Vlsr\ = $-$10 to 0 \kms\  in cyan and +21 to +34 \kms\ in red in the region containing the north and northwest fingers.  The yellow circles mark compact (unresolved) 1.3 mm continuum sources.  Some can be identified with stars in visual or near-IR images; others, such as SMA-15 \citep{Teixeira2016} have no visual or near-IR counterparts.   The outflows unrelated to OMC1 are indicated.  SMA-15 drives a bipolar flow with a redshfted lobe deflected towards the north.  The blueshifted limb-brightened cavity marks an outflow, possibly originating form SMA-14 in \citet{Teixeira2016}.  This flow appears to terminate in the compact knots just east of the label `SMA-14 outflow'.   These knots are associated with a southwest-facing bow shock in \Htwo\ that is also an HH object.  Finally, the highly collimated blueshifted filament labeled `SMA-18 outflow' traces a jet, possibly from SMA 18 in \citet{Teixeira2016}.  These flows are described in the text.
}
\label{fig10a}
\end{figure}

\begin{figure}
\begin{center}
\includegraphics[width=6in]{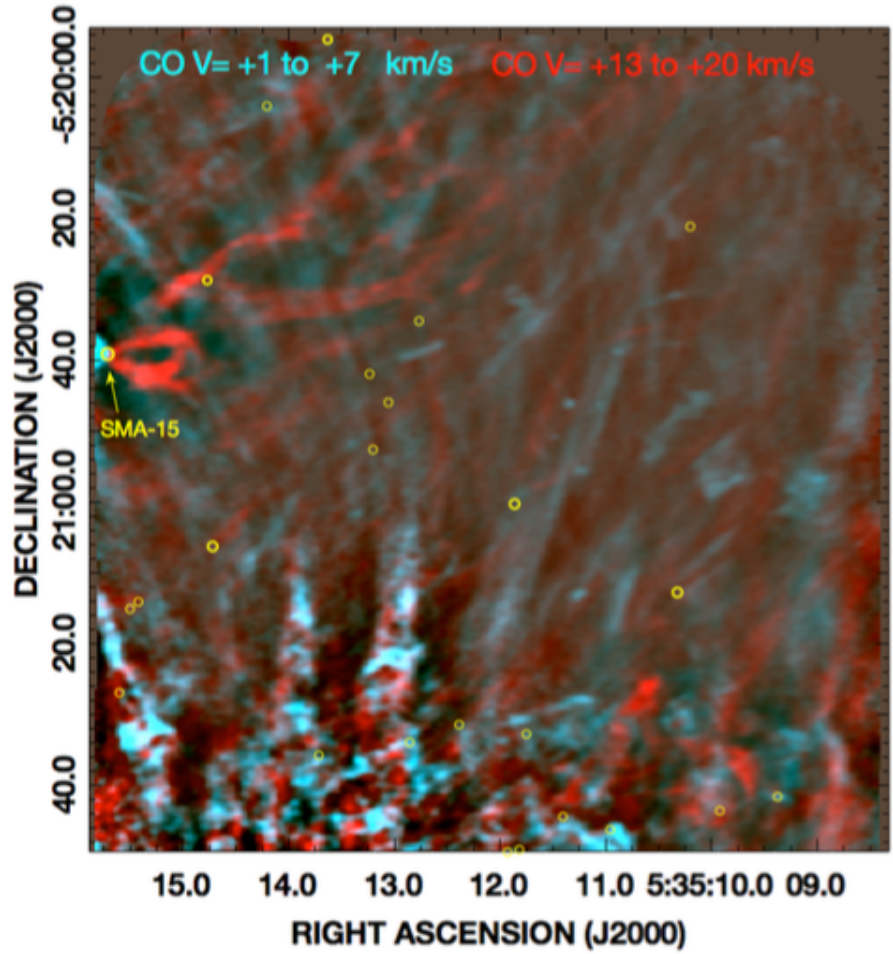}
\end{center}
\caption{An image showing the CO emission  in the region containing the north and northwest fingers showing emission at \Vlsr\ =  +1 to +7 \kms\ in cyan and +13 to +20 \kms\ in red.   Yellow circles are as in Figure \ref{fig10a}.
}
\label{fig10b}
\end{figure}

\begin{figure}
\begin{center}
\includegraphics[width=6in]{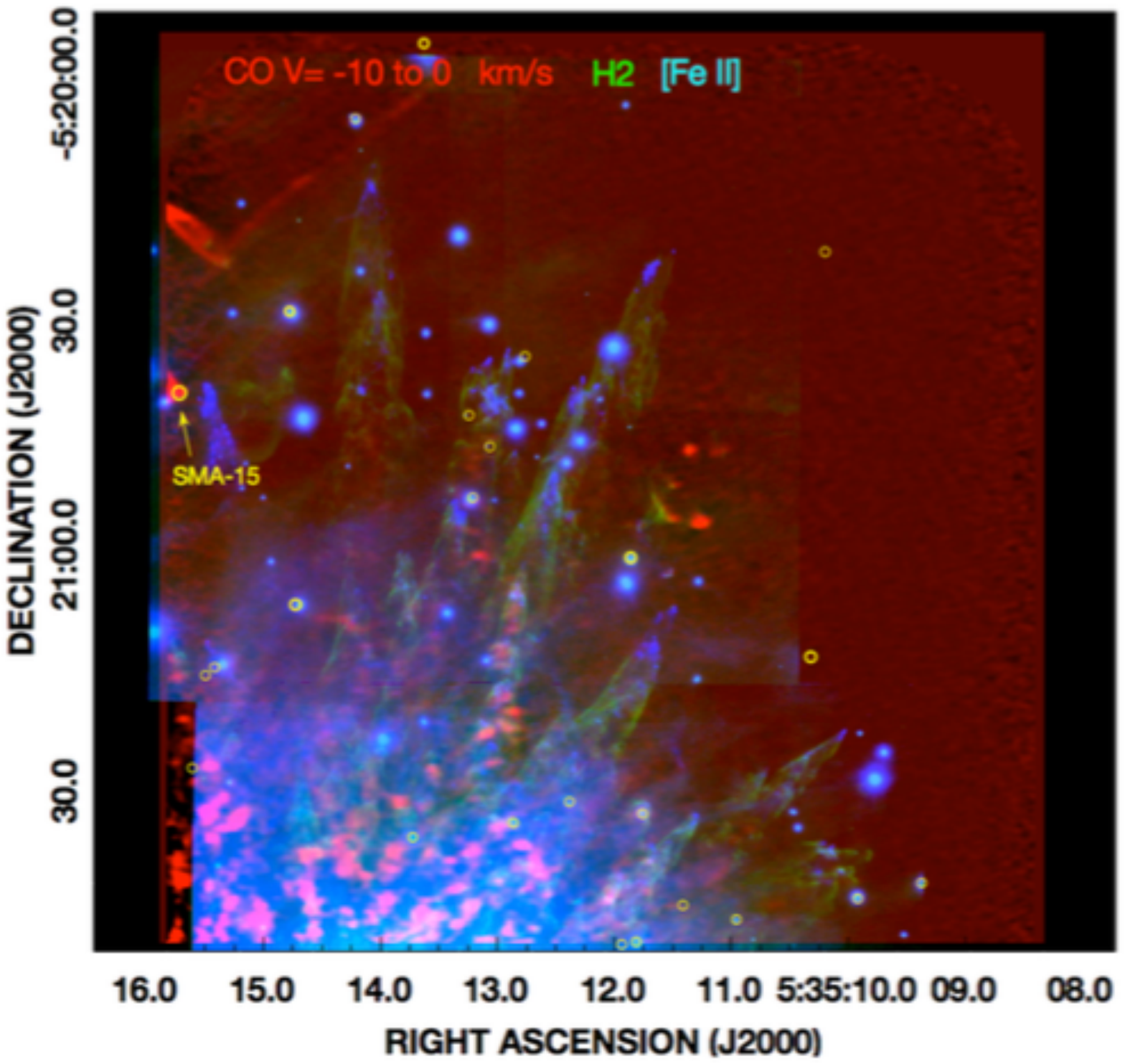}
\end{center}
\caption{An image showing the CO emission at \Vlsr\ = $-$10 to 0 \kms\  (red) superimposed on the \Htwo\ (green) and [\Feii ] (blue) emission from \citet{BallyGinsburg2015} in the region containing the north and northwest fingers.  
}
\label{fig10c}
\end{figure}

\begin{figure}
\begin{center}
\includegraphics[width=6in]{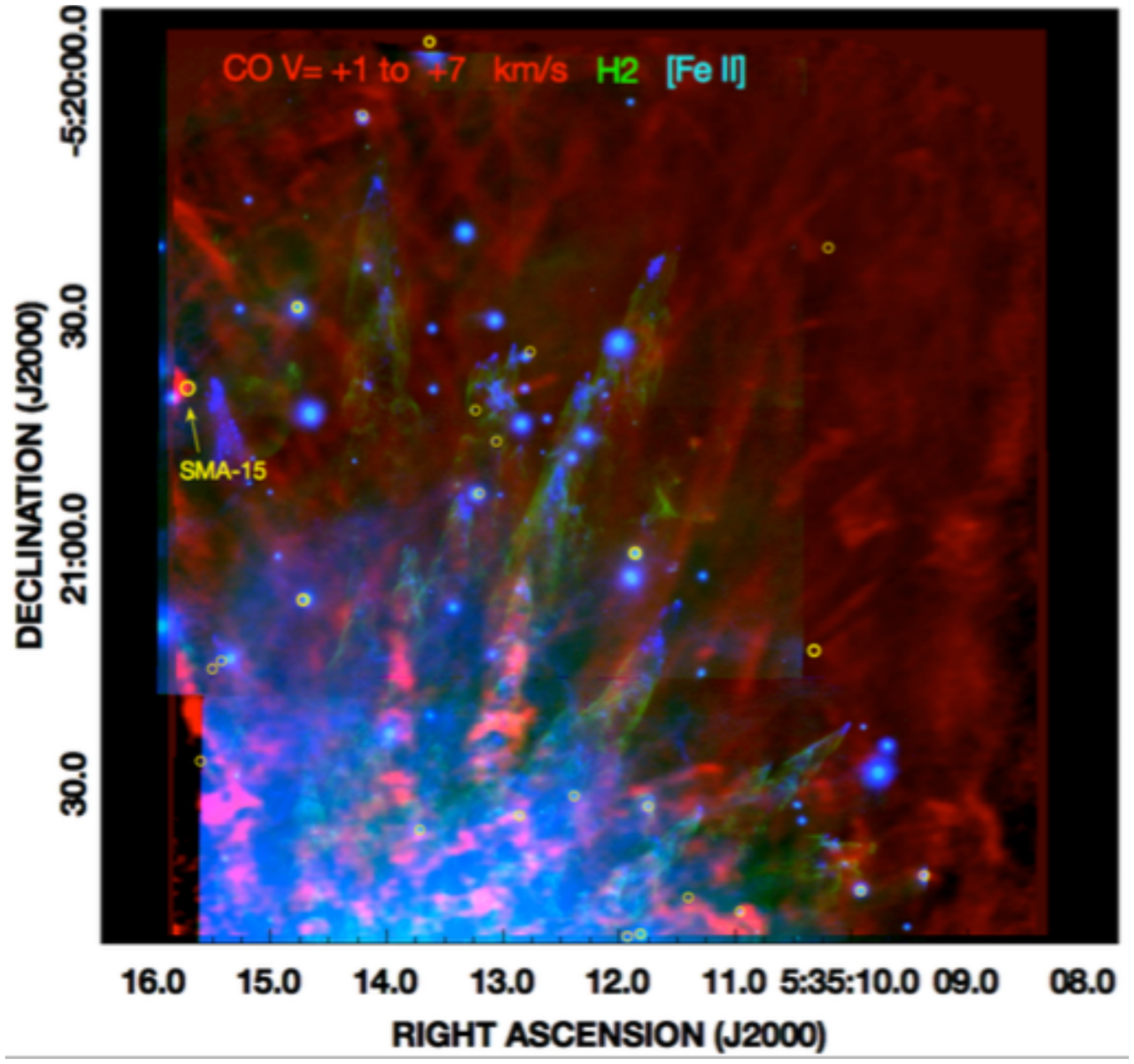}
\end{center}
\caption{An image showing the CO emission at \Vlsr\ = 1 to 7 \kms\  superimposed on the \Htwo\ and [\Feii ] emission from \citet{BallyGinsburg2015} in the region containing the north and northwest fingers.  
}
\label{fig10d}
\end{figure}

\begin{figure}
\begin{center}
\includegraphics[width=6in]{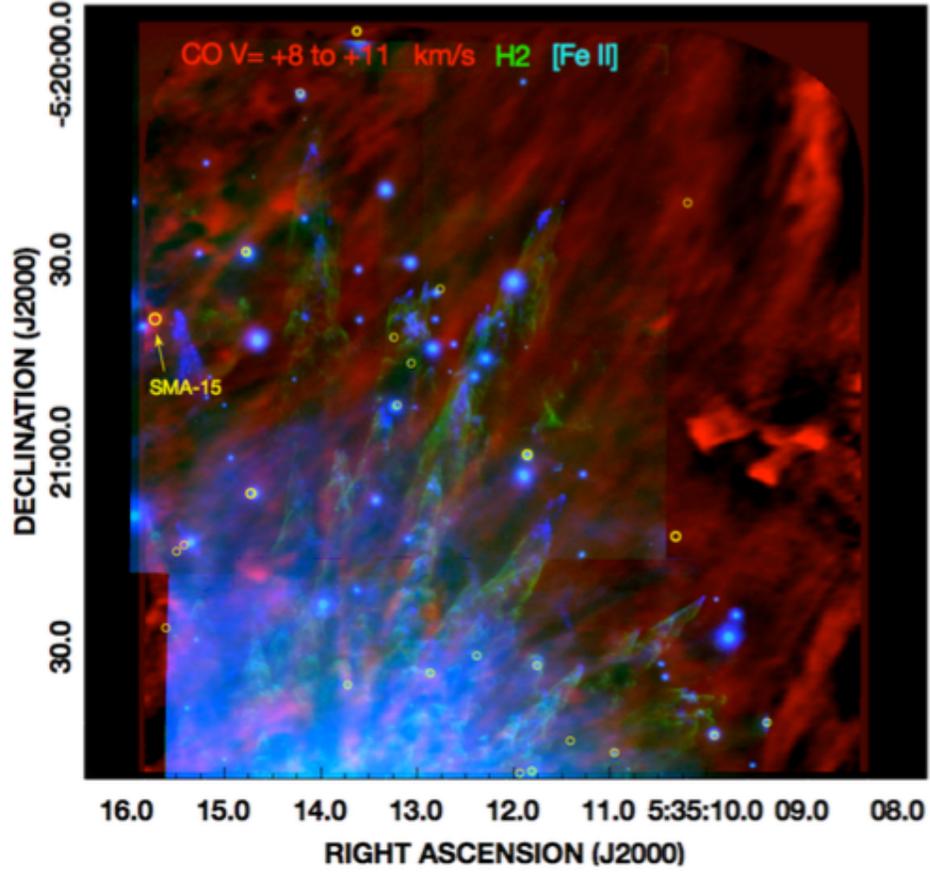}
\end{center}
\caption{An image showing the CO emission at \Vlsr\ = 8 to 11 \kms\  which corresponds to the line core, superimposed on the \Htwo\ and [\Feii ] emission from \citet{BallyGinsburg2015} in the region containing the north and northwest fingers.  
}
\label{fig10e}
\end{figure}

\begin{figure}
\begin{center}
\includegraphics[width=6in]{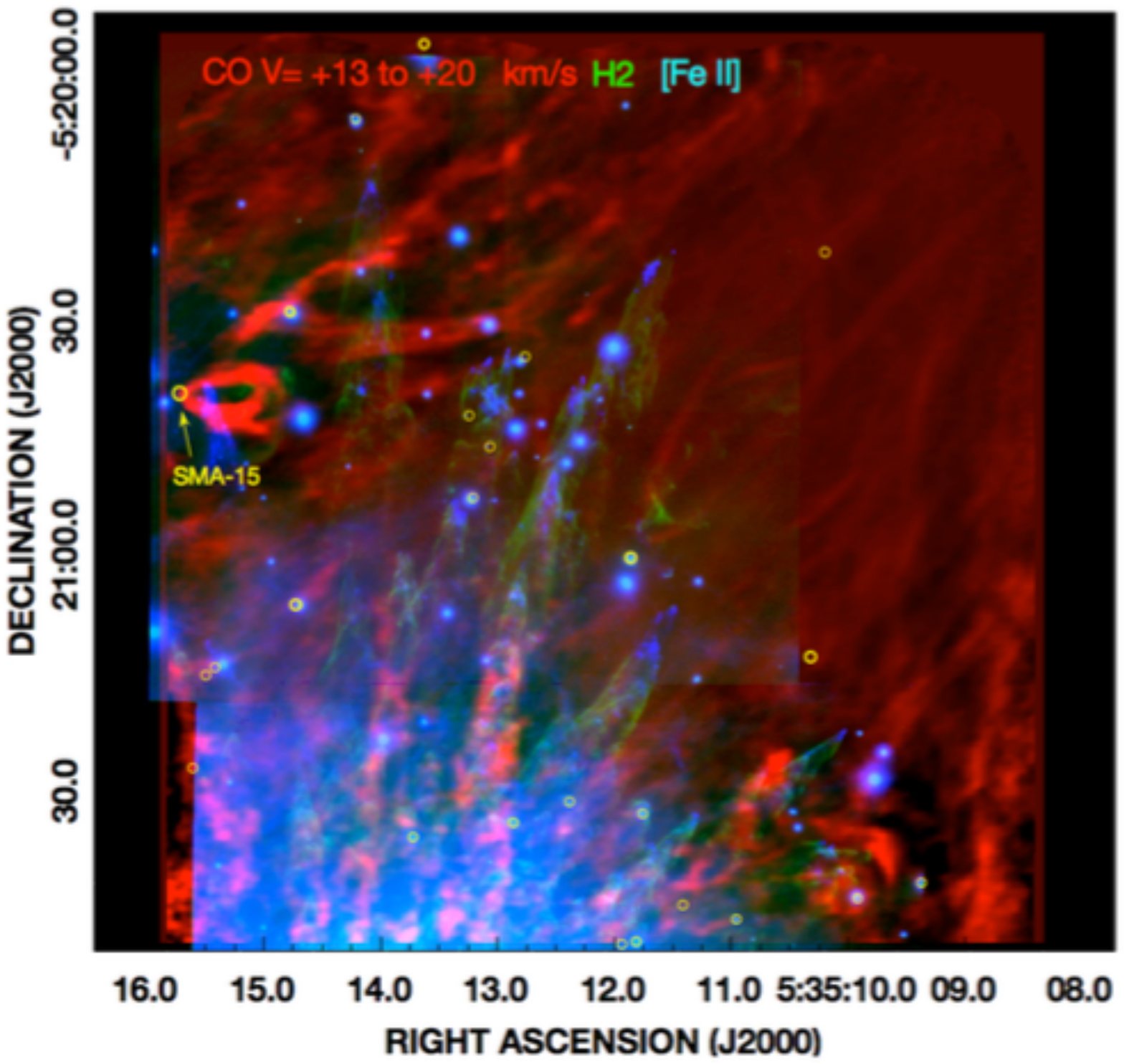}
\end{center}
\caption{An image showing the CO emission at \Vlsr\ = 13 to 20 \kms\  superimposed on the \Htwo\ and [\Feii ] emission from \citet{BallyGinsburg2015} in the region containing the north and northwest fingers.  
}
\label{fig10f}
\end{figure}

\begin{figure}
\begin{center}
\includegraphics[width=6in]{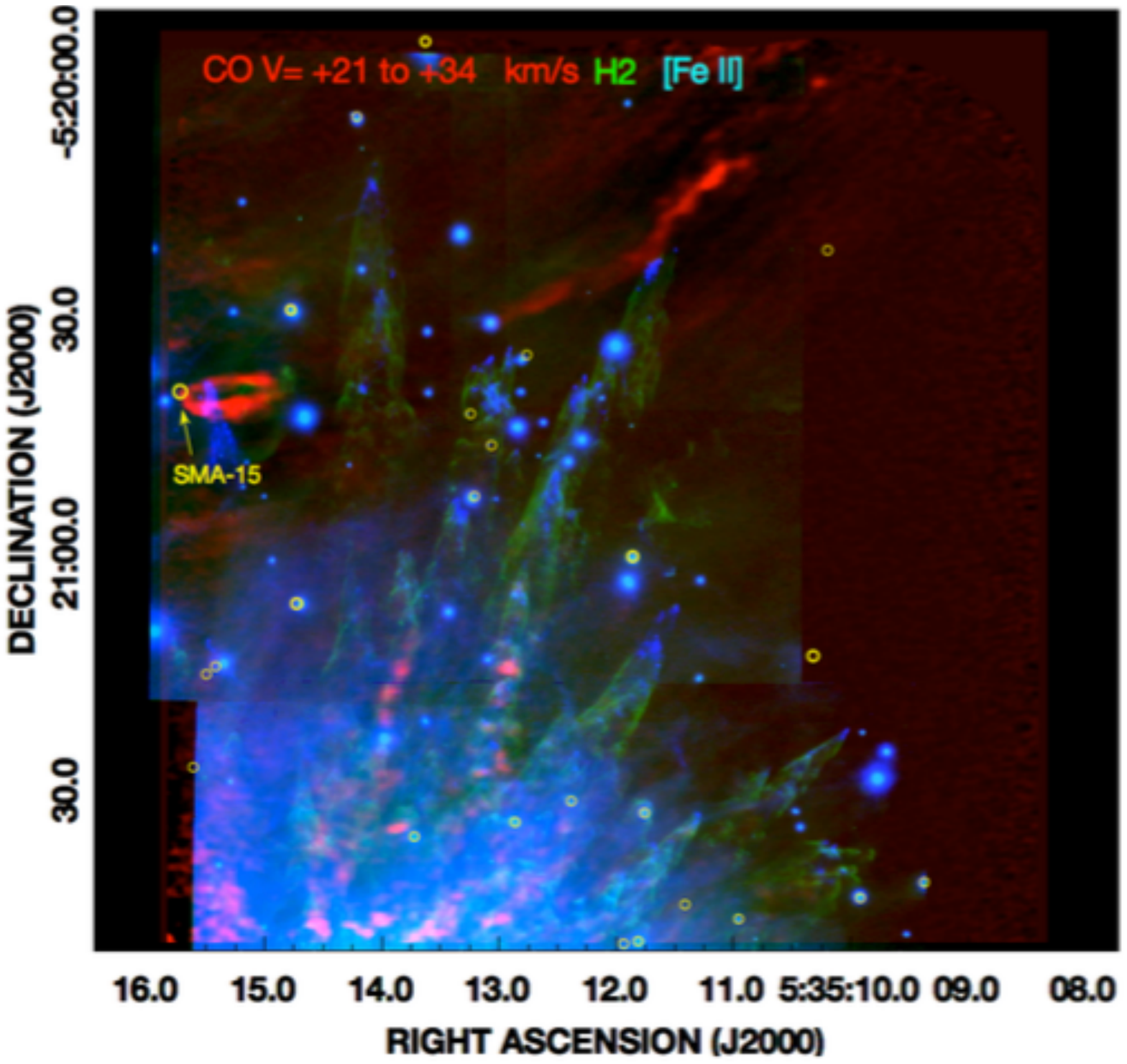}
\end{center}
\caption{An image showing the CO emission at \Vlsr\ = 21 to 34 \kms\  superimposed on the \Htwo\ and [\Feii ] emission from \citet{BallyGinsburg2015} in the region containing the north and northwest fingers.  
}
\label{fig10g}
\end{figure}

The largest CO Doppler shifts, $-$150 \kms\ and +145 \kms\  are located within $\sim$60\arcsec\ of the ejection center  (Table 1 and Figure \ref{fig4}). These line-of-sight motions  are about a factor of two smaller than the highest proper motions in \Htwo ,  [\Feii ],  and associated HH objects, 200 to 400 \kms\   \citep{Doi2002} located 100\arcsec\ to 140\arcsec\  north and northwest of the ejection center  (see Figures \ref{fig10a} through \ref{fig10g}).    The terminal radial velocities of most CO streamers confined to the inner $\sim$1\arcmin\ of the ejection center, are less than 100 \kms\, similar to the   \Htwo\ knot proper motions \citep{Bally2011}  and to the range of radial velocities where the brightest near-IR emission is located \citep{Oh2016}.   The factor of two larger motions in the  north and northwest \Htwo\ and [\Feii ] fingers and HH objects suggests that a fan of unusually fast ejecta was launched along the ejection direction of BN.

Figures \ref{fig10a} through \ref{fig10g} show that the northern  \Htwo\ fingers lack CO emission.  It is possible that some CO close to the radial velocity of Orion A is obscured by the foreground cloud.   Furthermore, in the presence of bright CO emission near the rest velocity of Orion A  and within tens of arc-seconds of OMC1 where emission extends over a wide range of velocities, residual beam artifacts from incomplete cleaning of the data and missing flux resolved out by the interferometer may hide emission below 10 K.  However, towards the northern \Htwo\ fingers, the noise in the spectral-line maps is less than 5 mJy/beam (away from the $\sim$7 to 12 \kms\  radial velocity range where CO emission from Orion A is bright).  Emission has to be fainter than 0.1 K to be below our sensitivity. 

Figure \ref{fig6} shows that \Htwo\ streamer \#1  is aimed directly towards HH~210 which is dominated by [\Feii ] emission with very little associated \Htwo .  HH~210 exhibits the highest proper motions, 380 to 425 \kms  \citep{Doi2002},  and contains $\sim$10 mega-Kelvin plasma observed by the Chandra X-ray telescope \citep{Grosso2006}.  HH~210 has a radial velocity of $-$64 \kms\  \citep{Doi2004} implying that it moves close to the place of the sky.  There are no CO features at any radial velocity that convincingly point toward this high-velocity shock from OMC1.   

HH~201, one of the brightest shocks associated with the OMC1 explosion at visual wavelengths, and along with HH~210, among the two brightest [\Feii ] dominated shocks in the Orion Nebula,  also has no clear CO counterpart.  \citet{Doi2004} found a radial velocity of $-$260 to $-$284  \kms\  in H$\alpha$ for this object, implying that it is moving out of the cloud towards us and into the Orion Nebula.   HH~201 lies in the region of the northwest-facing bubble seen at both red- and blueshifted CO radial velocities.   There is some low velocity CO at blueshifted radial velocities with respect to OMC1 at \Vlsr\ $\approx$ 2.4 to about 5.3 \kms\ in the northwest bubble several arcseconds beyond the tip of HH~201.

Although no CO streamers reach the ends of the north and northwest fingers, the bases of several of the most prominent \Htwo\ fingers contain clumpy, low-velocity CO with peak brightness temperatures ranging from 10 K to less than 0.5 K.    Figure \ref{fig10a} shows the summed emission from \Vlsr\ = $-$10 to 0 \kms\ at blueshifted radial velocities (in cyan) superimposed on  the redshifted emission at \Vlsr\ = +21 to 34 \kms\ (in red).  Clumps of CO  at both red- and blueshifted radial velocities are associated with fingers \#2, \#3, \#4, and \#5 (Figure \ref{fig6}).    Figure \ref{fig10b} shows the summed emission from \Vlsr\ = +1  to +7 \kms\  at blueshifted radial velocities (in cyan) with respect to the OMC1 rest frame, superimposed on  the redshifted emission at \Vlsr\ = +13 to +20 \kms\ (in red).    In this figure, only  streamers \#3 and \#5 are traced by CO.  The  widths of the CO emission orthogonal to the \Htwo\ features are larger than in the previous figure.    Figures \ref{fig10c}, \ref{fig10d}, \ref{fig10e}, \ref{fig10f}, and \ref{fig10g} show progressively higher velocity  CO emission integrated over the radial velocities indicated in the Figures from \Vlsr\ = $-$10 \kms\ to +34 \kms .    These figures suggest that the longest north and northwest  \Htwo\ fingers most likely lie in the plane of the sky.  

Figures  \ref{fig10a} and \ref{fig10b} show CO clumps associated with a CO streamer aimed east of streamer \#1.    This feature is only associated with very faint \Htwo\  and [\Feii ] emission with the brightest feature located at the east edge of the mapped field at   05:35:15.73,$-$5:21:12.6 at the left edge of Figures \ref{fig10c} to \ref{fig10f} about one-third of the way from the bottom.  This streamer and its associated shocked emission may be partially hidden by the dust ridge extending north-northeast from OMC1 towards OMC2. 
  
Several models might explain the gaps between the ends of the CO streamers and the near-IR shocks in the fingertips.   First, above the velocity of order $\sim$100  \kms\  observed  at the leading edges of many streamers, shocks may completely destroy CO.   Second, the fastest ejecta may have been launched in mostly atomic or ionized form and may not have had sufficient time, or the conditions, to form  molecules.  If either model is correct,  the gap between the fingertips and the CO streamer ends is expected to be filled with dense atomic gas or plasma; tracers such as sub-mm CI,  far-IR C$^+$,  or radio continuum should detect it.  Third,  the  gap may be empty, with dense clumps located close to the fingertips that are required to drive the observed high-proper motion shocks.     Future observations will distinguish between these scenarios.   Figure \ref{fig9} shows a cartoon illustrating the major observed features of the OMC1 outflow, the relative locations and motions of the ejected stars, and the suspected orientations of their disks. 

\subsection{Other Outflows in the Field}

Several protostellar outflows unrelated to OMC1 are also seen in the CO data.     \citet{Teixeira2016}  surveyed the field northeast of the region covered by our observations in the 1.3 mm bands.  They found a number of compact mm continuum sources and molecular outflows.    SMA-15 appears at the left edge in our maps and drives a prominent bubble and bipolar high-velocity outflow launched in an east-west direction.  The western, redshifted lobe of this flow overlaps with HH~210.  A small west-facing \Htwo\ bow-shock is located along the axis of this flow about 10\arcsec\ west of SMA-15.  A complex, redshifted outflow lobe can be traced for nearly 90\arcsec\ to the west and northwest and  appears to be deflected in this direction.   

\citet{Teixeira2016}  SMA-14, located just beyond out mapped field about 25\arcsec\ north of SMA-15  drives a highly collimated, blueshifted CO jet at PA $\approx$ 225\arcdeg .  In Figure \ref{fig10a} this flow can be seen as a 10\arcsec -long  cavity with a width of about 5\arcsec .  This flow is pointed slightly south of a collection of CO knots and an \Htwo\ bow facing towards the southwest.   If the collection of CO knots and the \Htwo\ bow are powered by the SMA-14 outflow, then this flow must  also experience a small northward deflection.  

A blueshifted, southeast-northwest oriented filament in the upper-left corner of Figure~\ref{fig10a} traces back to \citet{Teixeira2016}  SMA-18 or possibly SMA-17 beyond the east-edge of our data.    This feature may trace a highly collimated jet.   

Finally, in a small blueshifted outflow emerging from the OMC1-S region ($\sim$90\arcsec\ south of OMC1) is seen in the lower-right corner of Figure \ref{fig1b}.  This flow appears to be a collimated, knotty jet between \Vlsr\ $\approx -$8 to +4 \kms .  This flow may be related to HH~625 \citep{ODell2015,Kaifu2000}.
 
\section{Discussion}

Figures \ref{fig1} through \ref{fig3} show that the high-velocity CO streamers point back to a location between the three ejected stars.   Over 80\% of the streamer orientations trace back to within an elliptical region with major and minor axis diameters of $\sim 3 \times 5$ \arcsec\ oriented at PA $\sim$330\arcdeg\ and centered at 05:35:14.33, $-$5:22:27.8, shown as a cross in the Figures and close to the suspected ejection site of the OMC1 runaway stars when the proper motions are corrected to the reference frame of OMC1 as discussed below.  The coincidence between the streamer orientations and the site of stellar ejection indicates that the explosion in the gas was likely powered by the stellar ejection.    As discussed below, the energetics requires an AU-scale encounter between a pair of 10 \Msol\ stars,  or a sub-AU-scale encounter between less massive stars.    The fastest ejecta  can be created by the gravitational slingshot of circumstellar or stellar photospheric material as three or more stars approach each other at speeds of order the Kepler speed given by their masses and separations.   Dense gas in  OMC1 may  have re-shaped the fast ejecta into the observed Hubble flow streams as discussed in the next section.   Following ejection of the stars, orbital motion of material bound to the cluster prior to ejection would be converted into linear expansion to produce low velocity ejecta.   Numerical simulations of collisions or near misses show that debris is likely to be preferentially ejected along the orbital planes of the interacting stars during the close encounter \citep{PortegiesZwart2016}.     The bubbles and fans of low-velocity ejecta noted above may have their origins in such streams.   The masses and momenta of the ejected stars  are critical parameters that must be considered by any model of the OMC1 explosion.

\subsection{The streamers}

The high-velocity CO in OMC1 consists of a nearly isotropic distribution of over 100 linear streamers whose radial velocities increase with projected distance from the source.    Towards the northeast and southwest, the CO streamers and the distance of the \Htwo\  and [\Feii ] shocks in the Orion fingers have comparable spatial extent.  However, towards, the north and northwest, the distance of the CO streamers from the ejection region is about one half of the distance of the most distant  \Htwo\ fingers and  [\Feii ] fingertips.  The largest  CO radial velocities are also about two times smaller than the fastest measured proper motions of the \Htwo\  and [\Feii ]  shocks.   The CO morphology resembles the jet-like features seen in high-speed videos of powerful terrestrial explosions occurring inside a damping medium.   Experiments with explosives embedded in the center of a sphere of solid beads, wet-beads, or liquids produce hundreds of jet-like streamers with ejection speeds increasing with distance and a morphology similar to those seen in Orion \citep{Frost2012,Milne2016} (for videos, see  https://www.youtube.com/watch?v=78l4cU5F2YE).   Particle jets displaying a Hubble-law velocity-distance relation are a general characteristic of explosions occurring in an inert damping medium.

\citet{Stone1995} proposed that bullets responsible for driving the multiple \Htwo\ fingers and  [\Feii ] fingertips in the OMC1 outflow were produced by Rayleigh-Taylor (RT) hydrodynamical instabilities resulting from the interaction of a wind with the surrounding medium.  A swept-up shell is subject to RT instabilities if it accelerates;  it can be triggered by a wind with increasing mechanical luminosity,  a steady wind running into a medium with a steeper than inverse square-law density profile, or a combination of increasing wind power and a shallower density profile.    RT instabilities can form dense clumps whose wakes create low-excitation \Htwo\ fingers and high-excitation [\Feii ] fingertips.  \citet{McCaughreanMacLow1997} proposed a variant of this model in which multiple sources in a dense cluster lead to the formation of a fragmented shell.  

Ejecta from an explosion occurring inside a cocoon of relatively stationary material will be decelerated by interactions with the medium.  While the cores of sufficiently dense or massive clumps may retain much of their initial velocities as they move through the medium and exit its outer boundary,  the motions of less massive or lower density clumps will be damped.   Furthermore, the clump surface layers will be shorn-off, mixed with the medium, decelerated, and subjected to Kelvin-Helmholtz (KH) shear instabilities.     Assuming that the cocoon has an outer boundary,  the passage of sufficiently massive ejecta through the medium will result in the emergence of a stream of debris having a distribution of masses and a wide range of velocities.   In the absence of further deceleration, the ejecta will be sorted by velocity upon exiting the damping medium of the cocoon.  A finite time later, the distance travelled will be  proportional to the speed of ejecta as it leaves the  cocoon, forming a Hubble flow as observed in the OMC1 explosion.     

\subsection{The Masses of the OMC1 Runaway Stars}

Linking the OMC1 explosion to the dynamic ejection of its runaway stars requires knowledge of the stellar masses.   The mass of  the fastest ejected star, BN, has been estimated to be between 12 and 15 \Msol\ \citep{Scoville1983} or 8 to 10  \Msol\ \citep{Tan2004,ChatterjeeTan2012}.    Assuming that Source I has an oppositely directed momentum, it  must have a mass of  16 to 25 \Msol\    \citep{Goddi2011}.    Source n, also involved in the dynamic interaction,  has a mass of $\sim$3 to 6 \Msol .   Thus, to conserve momentum in a 3-way interaction involving Source n, Source I should have to have a mass of $\sim$10 to 22 \Msol\  to balance the momentum of BN.

Millimeter and sub-millimeter interferometric measurements of the rotation curve of the 50 AU radius disk surrounding source I indicate a  mass between only 5 to 7 \Msol\  \citep{Matthews2010,Plambeck2016}.   If this estimate is correct, then Source I plus n by themselves could not have been responsible for launching the BN object with a velocity of $\sim$26 \kms\ and the model proposed by  \citet{Tan2004}  in which BN was ejected from the Trapezium $\sim$4,000 year ago may be  more viable.   However,  the ejection of BN by the Trapezium leaves unexplained the 10 to 14 \kms\ motions of Source I towards the southeast and motion of Source n from the ejection center of the Orion fingers and CO streamers in a direction opposite to the motion of BN.    Furthermore,  a passage of BN to OMC1 sufficiently close to trigger an explosion is unlikely.   Using the projected separation of $\theta ^1 Ori$~C, the most massive member of the Trapezium group of massive stars in the core of the Orion Nebula,  and BN of 60\arcsec ($\sim$25,000 AU),  an encounter within a radius of 50 AU of source I has a probability of $\sim 10^{-6}$, assuming an isotropic  probability distribution of the ejection direction of BN.   However,   \citet{ChatterjeeTan2012} argue that such an event is not impossible and that the passage of BN through OMC1 could have triggered the explosion observed in the gas  by driving a major accretion event onto one of the OMC1 massive stars.       

As discussed by \citet{Matthews2010} and \citet{Plambeck2016},  it is possible that the Source I disk rotation curve underestimates the central star's mass because of internal pressure support.   If magnetic fields or turbulence is sufficiently strong, the disk would rotate at sub-Keplerian velocities \citep{Shu2008}.    Additionally, a strong outflow driven by magneto-centrifugal acceleration  exerts a strong braking torque on the disk surface.  Thus, the observed velocities may underestimate the midplane orbital velocities, and therefore the mass of the central star.

\citet{Plambeck2016} found that the Source I disk has a mass of 0.02 to 0.2 \Msol\ and a projected thickness of about 20 AU.  Using the 50 AU radius, these masses imply that the mean density of \Htwo\ ranges from $1.7$ to $17 \times 10^{10}$~cm$^{-3}$.    \citet{Girart2004} and \citet{Plambeck2003} found evidence for a magnetic field in source I.  Using the \citet{Crutcher2012} scaling of magnetic field strength  with density, $B \sim 10 (n  / 300 ~cm^{-3})^{2/3}$ micro-Gauss, where $n$ is the \Htwo\ volume density, the magnetic field strength in the Source I disk could be around 1 to 10 Gauss for the above densities.    The total magnetic energy of the Source I disk would then be $E_B \approx (B^2 / 8 \pi) \pi R^2_{disk} Z_{disk} \sim 2 \times 10^{43}$ to  $2 \times 10^{45}$ erg.   Here, $R_{disk}$ is the observed disk outer radius and  $Z_{disk}$ is its vertical extent.  The kinetic energy in orbital motion of the 0.02 to 0.2 \Msol\ disk with a characteristic orbit speed of 10 \kms\ is about $10^{43}$ to $10^{45}$ ergs, depending on the mass distribution.  Because the magnetic energy and disk orbital kinetic energy may be comparable,  magnetic breaking and  sub-Keplerian disk rotation is plausible.   Additionally, the young, thermal SiO and H$_2$O maser flow emerging from source I contains the brightest and highest density portion of the OMC1 outflow.   Thus it is likely to exert a substantial torque on the Source I disk:  The rotation-curve-based mass estimates for source I may be lower bounds.

A recent dynamical interaction and abrupt acceleration of Source I  to a velocity of $V_* \sim $13  \kms\ would have disrupted any pre-existing outer disk beyond the gravitational radius,  $R_G \approx GM / V^2_*$ = 148 AU  for  M= 22 \Msol , or 34 AU for M =  5 \Msol .     Disk material closer to the star than $R_G$ could remain gravitationally bound following acceleration by the dynamical interaction.    However, as discussed below, acceleration to these velocities requires an AU-scale encounter with the other stars.  Such a penetrating encounter would eject most of the disk beyond $R_G$ and would severely perturb the remainder inside this radius.   A deep, penetrating encounter leading to the acceleration of the interaction-formed binary or merger remnant would likely lead to a re-orientation of the  angular momentum vector of the surviving disk \citep{MoeckelBally2007a,MoeckelGoddi2012}.   The damping-time would be a few times the orbit time at the $\sim$50 AU outer radius of retained disk material, or longer than 75 to 160 years for M = 22 to 5 \Msol\ stars, respectively.   Material ejected by such close encounters that did not reach escape speed could be falling-back onto the remnant disk for many times the orbit speed at the disk outer radius, producing longer-lasting disk perturbations.   Additionally, if the interaction resulted in the formation of a compact binary, the orbits are likely to be highly eccentric, and non-coplanar with the surviving disk.    Gravitational forces from inclined, eccentric orbital motion would stir the disk and drive turbulence.  

Thus, it is possible that torques exerted by a powerful outflow, magnetic fields, and internal turbulence, contribute to sub-Keplerian rotation of the Source I disk, leading to an underestimate of the mass of Source I.

\subsection{The potential role of Source n} 

The fast proper motion of BN and the slow-motion of Source I could be partially reconciled with the recent measurement of the motion of Source n by \citet{Rodriguez2016}.    The momentum and kinetic energy in the motion of BN could be balanced by slower and lower-mass Source I plus the momentum and energy of Source n.  Source n has a luminosity of $\sim$2,000 \Lsol  ,  an estimated mass of 3 to 6 \Msol , and is surrounded by a  340 by  230 AU diameter disk detected at 8 and 11.7 $\mu$m \citep{Greenhill2004}.  The Source n disk emits in the 2.3 $\mu$m overtone bands of CO \citep{Luhman2000}.  A pair of radio sources and a jet-like feature extending to the southwest  at PA $\sim$200\arcdeg\  indicate that it may be a binary YSO  which powers a one-sided jet roughly orthogonal to the disk  \citep{Rodriguez2016}.  Recently \citet{Rodriguez2016}  detected  a 15.4 \kms\  proper motion of the compact binary core of the radio image towards the south.    All three of the OMC1 radio sources,  Source n, BN, and Source I  were within 110 AU of each other in the year AD 1475$\pm$6 \citep{Rodriguez2016}.  Source n, along with BN and Source I  may have played an important role in  the dynamics of the OMC1 about 540 years ago.   

A proper motion of  $\sim$15  \kms\ implies that the trajectory of Source n would have intersected the trajectory of source I between 370 to 540 years ago;   if the proper motion were 26 \kms\ as suggested by \citet{Gomez2008}, the intersection would have occurred 270 and 345 years ago.   In the model in which  a dynamical decay  launched source I and BN $\sim$540 years ago,  Source n  was either ejected during this interaction,  several hundred years after the main event, or it is yet another moving star whose origin is unrelated to the event that launched the massive stars.   It is possible that the event 500 years ago involved 4 or more stars in which the interaction produced an unstable triple or quadruple moving in the opposite direction to BN.   This system could have  decayed up to a few hundred years later and launched Source I and Source n on their current trajectories.    Such a staged disintegration could explain the several arc-second dispersion in the streamer orientations.    In this model either Source I, Source n, or both would have to be a compact binary or merger remnant \citep{ReipurthMikkola2012,Reipurth2015}.  

The  sum of the 5 to 7 \Msol\ mass of Source~I from \citep{Plambeck2016}  and the  3 to 6 \Msol\ mass of Source n is still at most 13 \Msol  .  Assuming these masses, and using the stellar speeds in our assumed OMC1 reference frame, and adding the momenta of these two ejected stars  give a total momentum of 98 to 164 \Msol ~\kms , a factor of nearly two to three less than the 230 to 430 \Msol ~\kms\ momentum of BN.     No reasonable shift of the OMC1 rest frame can bring these momenta into agreement.   Thus, either the mass of Source I or n has been severely underestimated,  or there is another ejected massive member that has not yet been identified.    Other massive objects that might have participated in the interaction include other stars, or high density and massive compact cores  such as IRc4.  

\subsection{The energetics of the OMC1 explosion: }
 
It is possible that the OMC1 core has a net motion of few \kms\ with respect to the stars in the Trapezium region.   Within 1\arcmin\ of OMC1, the radial velocity of molecular gas increases from about 8 to 11 \kms\ from south to north.  The gravitational potentials of the Trapezium cluster  and OMC1, each of which contains a mass of around 100 \Msol\ and are separated by about 0.1 pc,  could produce a relative motion of order 3 to 4 \kms , comparable to the observed change in radial velocity of the molecular tracers in the vicinity of OMC1.   The \citet{Dzib2016,Rodriguez2016} proper motions,  referenced to a frame in which 79  Orion radio sources are at rest, are shown as solid vectors in Figures \ref{fig1}, \ref{fig1b}, and \ref{fig3}.   These vectors intersect several arc-seconds east of the ejection center indicated by the earlier measurements and significantly displaced from the center of mass of the three stars assuming mass-ratios of 4:2:1 for Source I,  BN, and  Source n, respectively.   This location is also several arc-seconds east of the point where a traceback of the CO streamers intersect.     The dashed proper motion vectors in Figures \ref{fig1}, \ref{fig1b}, and \ref{fig3}  were derived by assuming that OMC1 is moving west with respect to the Trapezium stars with a velocity of 4 \kms  towards PA = 273\arcdeg .  Such a motion brings the ejection center to coincide with the center of mass of the stars assuming the above mass ratios.

The proper motions of BN, Source I, and Source n were discussed in the introduction.   
The radial velocity of Source n is unknown.  However,  given that it is visible in the near-IR,  it must suffer less extinction than the nearby Source I, so it may be on the nearside of  OMC1 and  likely blueshifted.    The lower speed of Source I, combined with 5\arcdeg\ to 10\arcdeg\ estimated error in the orientation of its proper motion may account for the differences in its  ejection direction measured by the various groups.      The proper motions of BN, Source I, and n in the OMC1 reference frame discussed above are  27, 12, and 17 \kms , respectively.   Combining the radial velocities with the proper motions, and assuming that the line-of-sight  velocity of Source n is $-$5 \kms , the total speed of Source BN, Source I, and n are about 27, 13, and 17 \kms .
 
If the masses of these stars are 8 to 16, 10 to 25, and 3 to 6  \msol , respectively, the sum of the kinetic energies in the stellar motions is $E_{star} \approx 1 - 2 \times 10^{47}$ erg.    The kinetic energy in the outflow today has been estimated to be $E_{outflow} \sim 4 \times 10^{46}$ \citep{Snell1984} to $4 \times 10^{47}$ erg \citep{KwanScoville1976}.    The energy source responsible for the ejection must supply at least $E_{tot} \approx 10^{47}$ erg during the ejection event.    However, much of the initial kinetic energy of the outflow would have been converted into radiation by shock waves that render the outflow visible in CO, \Htwo , and other species.   Accounting for radiative losses by shocks over the last 500 years, and the work required to climb out of a $\sim$400 AU region in the gravitational potential well of the $\sim$100 \Msol OMC1 cloud core, the energy source responsible for the ejection must supply at least $E_{tot}$ = $10^{48}$erg.
 
The kinetic energy of the runaway stars ejected by N-body interactions in compact groups is thought to come from the formation or hardening of  compact binary systems or stellar mergers.    The formation of a binary by a 3- or -4-body interaction, or the interaction of a pre-existing binary with other stars that leads to the hardening of the binary (shrinking of the semi-major axis) releases  gravitational potential energy,  $E_G \approx G M_1 M_2 / 2 R$ where R is the final semi-major axis of the  binary and  $M_1$ and $M_2$ are the stellar masses.  Assuming that Source I is such a binary \citep{Goddi2011}, and that  $M_1$ =  $M_2$ = 10 \Msol ,  implies that the final binary semi-major axis must be $R \sim G M_1 M_2 / E_{tot} < $  0.9 AU to release  $E_{tot} \approx10^{48}$erg.  

The fastest ejecta in the OMC1 outflow have speeds in excess of 400 \kms .  Although only a tiny fraction ($<<$1 \Msol ) of the total mass has such high speeds, these speeds constrain the distance of the launch point  from the source star under the assumption that the ejection velocity is comparable to the escape speed from the ejection region.   Assuming a 20 \Msol\ star (e.g. Source I), and no boosting of the velocities by forces other than gravity, the fastest ejecta must originate from a distance of order  $R <  3 \times 10^{12}$ cm ($\sim$ 0.2~AU) from the star's center of mass.  

\subsection{Radio Source I;  a bloated protostar: }

Massive stars are thought to accrete at high rates from their parent cloud cores.  To form a star with a mass of 10 \Msol\ in $10^5$ years requires an accretion rate $\dot M \sim 10^{-4}$ \Msol\  yr\per .     Protostars experiencing high accretion rates develop extended and cool photospheres \citep{HosokawaOmukai2009}.   Accretion rates of $\dot M \sim 10^{-4}$ to $10^{-3}$ \Msol\  yr\per\  result in photospheric radii reaching values between 0.2 to 0.7 AU when the star reaches masses of 5 to 8 \Msol.    Thus, accreting massive protostars are expected to resemble cool, red supergiants.
  
Light from  radio Source I  illuminates a near-infrared reflection nebula extending approximately 1' east of the OMC1 core.  Source I has been indirectly detected at near-IR wavelengths (3.8 and 4.7 $\mu$m)  through light scattered into our line of sight by this reflection nebula \citep{Sitarski2013}. The spectrum of this nebula  indicates that the illuminating source has an effective temperature $T_{eff} \approx $ 4,000~K, similar to a K or M supergiant \citep{Morino1998,Testi2010}.   At this effective temperature, the radius of the stellar photosphere is $R \sim 1.5 \times 10^{13} L^{1/2}_4 / T^2_{eff}$ cm $\approx$ 1 AU  where $L_4$ is the luminosity in units of $10^4$ \Lsol .   The luminosity of the OMC1 core is about $10^5$ \Lsol .   Subtracting the contribution from the two known luminous sources, the BN object ($L \sim 1 - 2 \times 10^4$ \Lsol ), and IRc4 ($L \sim 2 \times 10^4$ \Lsol ) sets an upper bound on the luminosity of radio Source I, of $L_I \sim 6 \times 10^4$ \Lsol .  Given the effective temperature of Source I, and its luminosity, one can estimate an upper bound on the photospheric radius, 
$R_I \approx ( L_I /  4 \pi \sigma T^4_{eff} )^{1/2} <  3.6 \times 10^{13}$ cm, or $R_I < 2.4$ AU.   

A lower bound on the photospheric radius can be obtained by assuming that the luminosity is entirely due to accretion.  Setting the photospheric luminosity equal to the accretion luminosity gives
$
R \sim (G M \dot M / 4 \pi \sigma T^4_{eff})^{1/2}
$
where $M$ is the mass of the protostar, $\dot M$ is the accretion rate, $G$ is Newton's gravitational constant, and $\sigma$ is the Stephan-Boltzmann constant.   For $T_{eff} = $4,000 K, accretion rates of $10^{-5}$, $10^{-4}$, and  $10^{-3}$  \Msol\  yr\per\  onto a 10 \Msol\ protostar give photospheric radii of $R \sim$  0.11, 0.24, 0.5 AU, respectively.     Note that in a hot core with a gas temperature of $T_{gas} \sim$ 400 K, the sound speed is about $c_s \sim$ 1 \kms\ in molecular gas.  The expected accretion rate (assuming the core is roughly an isothermal sphere) is $\dot M \sim c^3_s / G$ $\sim 3 \times 10^{-4}$ \Msol\  yr\per .    These upper and lower bounds imply that prior to the ejection event $\sim$500 years ago, massive protostars in the OMC1 core likely had AU-scale photospheres.   

If at least one of the stars involved in the interaction 500 years ago was accreting at such a high rate and was bloated, the compact binary configuration required by the energetics implies that the binary semi-major axis is smaller than the photospheric radius of that massive protostar.   The companion would have penetrated the bloated star's photosphere where dissipation could have led to a stellar merger.  

\subsection{An Interaction-formed binary or  protostellar merger?}

The kinetic energy observed  in the outflow plus the motion of the ejected stars requires the release of $\sim 10^{48}$ ergs.    If the interaction produced a compact binary consisting of a pair of $<$10 \Msol\  stars,  their current mean separation has to be $<$ 2 AU to release this much binding energy.    However, if the massive stars in such a binary had bloated photospheres,  a mean separation of less than 2 AU would have led to a stellar merger.   The stellar collision would occur with a velocity comparable to the escape speed from the more massive star's surface,  roughly $\sim$100 to 300 \kms .    Because the interacting stars are young, they are unlikely to be tidally locked as is expected to be the case for mature short-period binaries that merge such as V838 Mon \citep{SokerTylenda2006}.   Any circumstellar material and the outer  stellar atmosphere would be ejected by the collision of photospheres.  If a merger occurred, the bulk of the gravitational potential energy release will then occur when the dense stellar cores merge.    A 10 \Msol\ protostar can release more than $10^{48}$ erg required to explain the kinetic energy of stellar motions and the outflow in Orion if the merger partner has mass larger than about 0.03 \Msol .  Thus, even relatively low-mass merger partners can release sufficient gravitational potential energy to produce the OMC1 outflow.   

The disruption of circumstellar disks by close-in ($<$ 1 AU) stellar encounters would eject debris with speeds comparable to the Kepler speed ($>$ 100 \kms ).    Following removal of 30 to 40 \Msol\  of stellar mass from the OMC1 core by the ejection of its massive stars, any envelope or circumstellar material surrounding the system that was gravitationally bound to it would expand with speeds ranging from 3 to 30 \kms\  if originating from within 1,000 AU to 10 AU of the center of mass.    In a merger, the bulk of the gravitational potential energy is released by the hypersonic collision of the stellar photospheres followed by the in-spiral of the stellar cores.  

Scattered-light images, polarimetry, and direct imaging show that Source I and BN are both surrounded by disks whose major-axis dimensions on the sky are parallel to their respective proper-motion (PM) vectors \citep{Reid2007,Jiang2005}.  Thus the disk spin-axes are at nearly right angles to the PM-vectors.    A disk orientation with a spin-axis approximately  perpendicular to the velocity vector of an ejected star is expected from angular momentum conservation by fall-back of debris that failed to reach escape speed during the ejection event, or  material accreted from the environment by the Bondi-Hoyle process  \citep{Bally2011,MoeckelGoddi2012}.   

Massive stars with near-Solar abundances of the heavy elements, even ones still forming, fuse hydrogen in their cores.    If the stars involved in a merger have masses of more than a few  \Msol , there could be a nuclear component to energy generation.   By the time a star reaches this mass, its core will be fusing hydrogen, mostly by means of the CNO Cycle which has an incredibly steep temperature dependence with $\dot E_{CNO} \propto T^{16}_{core}$\citep{Prialnik2000}.  Thus, it is possible that in a stellar merger,  perturbations to the pressure and temperature of the core could lead to a dramatic increase in the core luminosity which could add to the energy of the stellar envelope.    Modeling is needed to see how such an event would impact the stellar structure and properties such as its luminosity  and surface dynamics to determine if such en event can contribute to the explosion.

The Kelvin-Helmholtz (K-H) cooling time for a merger remnant, $L \sim G M^2 / RL = 1,400 ~M^2_{10} R^{-1}_{1} L^{-1}_4$ years where 
$M_{10}$ is the remnant mass in units of 10 \Msol ,
$R_1$ is its photospheric radius in units of 1 AU, and
$L_4$ is its luminosity in units of $10^4$ \Lsol .  Thus, a 1 AU radius $2 \times 10^4$ \Lsol\ object is expected to have a K-H cooling time of $\sim$700 years.
Such a remnant, formed about 500 years ago, would still be expected to resemble a supergiant star.  As discussed above, source I  resembles such a high-luminosity, cool object.  

\subsection{Explosive outflows and infrared-transients} 

Orion may not be unique.   Spitzer detected 4.5 $\mu$m emission from a wide-angle outflow emerging from  a high-luminosity ($\sim 10^6$ L$_{\odot}$) hot-core in  G34.26+0.15 in the inner Galaxy having a morphology similar to the \Htwo\ fingers in Orion \citep{Cyganowski2008}.  Unfortunately, this flow and its source  are at a distance of  $\sim$5 kpc and highly obscured.    Source G in W49,  the most luminous water maser outflow in the Milky Way, may be another example \citep{Smith2009}.   Interstellar bullets similar to the OMC1 fingers emerge from the massive protostar  IRAS 05506+2414 \citep{Sahai2008} and from DR21 \citep{Zapata2013}.   

In high-density, cluster-forming environments such as Orion where proto-stellar number densities can exceed $10^4$ to $10^5$ stars per cubic parsec, encounters between massive stars with bloated photospheres and other cluster members are likely to be relatively common.  Such encounters are facilitated by the dissipative nature of dense cloud cores which drain orbital angular momentum rapidly from the most massive protostars, causing them to be dragged into the center of the gravitational potential well of the star forming region.   The high multiplicity  among massive stars may be a consequence of such interactions \citep{MoeckelBally2007a} and may promote violent interactions between binaries and single stars or other binaries \citep{Goddi2011} that may lead to mergers.     Orion-like events may thus be relatively common in massive star forming complexes, occurring one or more times during the birth of a massive star.  Their shock-generated signatures would be erased on a time-scale required for the shocks to cross the dimensions of the cloud core,  $\sim 10^3$ to $\sim 10^4$ years.   However, such events may produce luminous infrared flares \citep{BallyZinnecker2005} and may be a major source of transient luminosity, kinetic energy, and momentum feedback in the self-regulation of star formation in regions forming dense star clusters and massive stars.

\citet{Smith2016} found a luminous visual transient, NGC~4490-OT, in the galaxy NGC~4490.  This object is  a candidate massive stellar merger event similar to but more luminous than V838 Mon.    The Spitzer space telescope during its warm mission phase found over 50 luminous transient events in galaxies located within 20 Mpc  at 3.6 and 4.5 $\mu$m  but with no counterparts in visual or near-IR images \citep{Jencson2016a,Jencson2016b,Kasliwal2016}.         Most are found in the dusty spiral arms of star-forming galaxies  such as M83.   These objects have luminosities intermediate between novae and supernovae.     Although some may be supernovae, ultra-luminous novae hidden  behind  dense clouds, or self-obscured within their own envelopes, the number of these luminous IR-transients suggests that they trace other types of eruptive phenomena such as stellar mergers in compact binary systems, eta-Carina type events, or Orion-like eruptions.

\section{Conclusions}

We present new observations of the Orion OMC1 outflow in the 230 GHz J=2$-$1 CO line with a resolution of $\sim$1 \arcsec .  The new data confirm the explosive nature of the outflow.   The main results of this study are:

$\bullet$   
The high-velocity CO emission consists of over 100 linear streamers of CO emission extending to radial velocities of $\pm$100 \kms . The streamers exhibit a roughly isotropic, spherically symmetric distribution exhibiting a `Hubble flow' with the radial velocity proportional to the projected distance from the explosion center.   The linear velocity-distance relation implies that the ejecta are orders of magnitude denser than the medium through which they move.   They tend to be brightest at their leading edges. 

$\bullet$ 
Towards the northeast and southwest, the CO streamers nearly reach the locations of the \Htwo\  and [\Feii ] shocks.  Furthermore, the range of CO radial velocities is comparable to the range of values of the  \Htwo\ proper motions  when the fastest  [\Feii ] features  and HH objects are excluded.

$\bullet$ 
Towards the north and northwest, the bright $T_B >$ 10 K CO streamers have a projected  spatial extent  about a factor of two smaller than the  extent of the outflow as traced by \Htwo\ and [\Feii ].  No CO clumps or bullets are seen at the leading edges of the north or northwest oriented fingers down to a brightness temperatures of order 0.1 K.  A few of the north and northwest fingers of \Htwo\  contain faint clumps of low-radial velocity CO along most of their lengths, many of which are associated with \Htwo\ and [\Feii ] bow shocks.  However, the two brightest [\Feii ] shocks which are associated with HH~201 and HH~210 have no clear CO streamers aimed at them.  The highest  red- and blueshifted radial velocities are about a factor of two  smaller than the fastest proper motions measured in these near-IR lines.   These features suggest that most of the ejecta towards the  north or northwest is moving very close to the plane of the sky.  However, it is also possible that shocks  dissociate accelerated molecular gas,  that the gas was launched as atomic or ionized material and did not form molecules, or there is a density gap between the shocks traced by near-IR emission lines and the ends of the CO streamers.    

$\bullet$
Several limb-brightened bubbles of ejecta emerge from the OMC1 region at velocities within tens of \kms\ of the line core.   The most prominent bubbles  are the redshifted feature extending to the east and the blueshifted bubble extending towards the southeast whose southern rim lies behind the Trapezium.    A less coherent bubble of both red- and blueshifted gas extends towards the northwest.

$\bullet$   
Deviations from spherical symmetry provide clues about the origin of the OMC1 explosion.   High velocity ejecta extend toward the northwest, and bubble-like features extending towards the east and southeast that may represent streams of dynamically ejected debris launched in a sub-AU-scale encounter of three or more stars.    The previously detected young, thermal SiO and H$_2$O maser flow emerging from source I contains the brightest and highest density portion of the OMC1 outflow.  

$\bullet$ 
The $\sim$500 year dynamical age of the outflow coincides with the time when radio Source I, the BN object, and Source n were closest to each other.    The coincidence between the age of the outflow and the time of closest approach of the stars  suggests that both the outflow and the stars were ejected from OMC1 by a dynamical interaction of four or more protostars.   It is also possible that the initial interaction with BN, produced an unstable system that decayed up to a few hundred years later and launched Source I and Source n on their current trajectories.  Such a staged disintegration could explain the several arc-second dispersion in streamer orientations.   

$\bullet$   
The energy of the event was likely powered by the dynamical interaction of a group of stars in the OMC1 core that led to either the formation of a compact (AU-scale) binary containing a pair of massive stars, or the dynamically induced merger of two stars.   Such an event may have produced an infrared-only  flare with a luminosity between novae and supernovae and  a duration of years to decades.

The OMC1 explosive outflow and stellar ejection poses many puzzles.  Is the mass of Source~I as low as  suggested by Plambeck and Wright (2016)?  If so,  how is the momentum of the BN object which is moving towards the northwest  balanced by stars ejected in the opposite direction?  Are there additional ejected stars, or compact masses such as nascent protostars like IRc4 involved?   Was the mass in the hot core or other compact clumps involved in the dynamic ejection?  Or, is the Source I disk in sub-Keplerian rotation, possibly because of internal pressure support provided by magnetic fields, turbulence, or perturbations associated fall-back and dynamic ejection a half-millennium ago?   How were the hundreds of CO streamers produced?   The CO streamers do not reach the most distant HH objects, \Htwo , and [\Feii ] shocks in the OMC1 BN/KL  outflow.  What continues to drive the fast motions of these shocks?  How common are OMC1-like explosive outflows?  How much do such events contribute to feedback and self-regulation of star formation?

\acknowledgments
This work was  supported in part by National Science Foundation (NSF) grant AST-1009847.
This paper uses ALMA data obtained with program
ADS/JAO.ALMA \#2013.1.00546.S. ALMA is a partnership of the European Southern Observatory
(ESO) representing member states, Associated Universities Incorporated (AUI)
and the National radio Astronomy Observatories (NRAO) for the National Science Foundation
(NSF) in the USA, NINS in Japan, NRC in Canada, and NSC and ASIAA in Taiwan,
in cooperation with the Republic of Chile. The Joint ALMA Observatory (JAO) is operated
by ESO (Europe), AUI/NRAO (USA), and NAOJ (Japan).
The National Radio Astronomy Observatory is a facility of the National Science Foundation operated under cooperative agreement by Associated Universities, Inc.
L.A.Z. is grateful to CONACyT, Mexico, and DGAPA, UNAM for their financial support.
J.B. thanks Professor Andreas Burkert from the Ludwig-Maximilians Universitat Munchen,  Department fur Physik,
for insightful discussion and for producing numerical simulations of streamer formation and Professor
Luis F. Rodrigues for sharing the results of recent radio proper motion measurements of
sources in OMC1.

\bibliography{new.ms}


%% for the paper.  Each keyword is check against the master list during
%% copy editing.  Individual instruments can be provided in parentheses,
%% after the keyword, but they are not verified.

\vspace{5mm}
\facilities{ALMA}

\software{IRAF, ds9, CASA \citep{casa}, RADEX \citep{vanderTak2007}}

\clearpage

\begin{table}
\caption{Highest Radial Velocity CO Knots in Orion OMC1}
\vspace{0.1in}
\small
\begin{center}
\begin{tabular}{l l l l r l }
\hline \hline
 No. & $\alpha$(J2000)$^1$  	&  $\delta$(J2000)$^2$ 	&  \Vlsr  $^3$& D$^4$  & Direction  \\ 
        &  (05$^h$35$^s$)  	&           ($-$05$^o$)                  & (\kms )  & (")  &    \\
\hline
1  & 13.99  & $-$22:11.9     & $-$149.5  &  17.8  &  N    \\
2  & 14.76  & $-$22:31.6     & $-$148.0  &   6.4   &  E \\
3  & 13.00  & $-$22:15.8     & $-$140.7  &   24.0 &  NW   \\
4  & 14.07  & $-$22:06.7     & $-$140.7  &   22.6 &  N  \\
5  & 14.32  & $-$21:30.7     & $-$137.8  &   58.0 &  N    \\
6  & 13.81  & $-$21:56.6     & $-$136.3  &   33.2 & NNW   \\
7  & 14.50  & $-$22:36.9     & $-$130.5  &   8.5   &  S \\
8  & 15.05  & $-$22:20.7     & $-$130.5  &   13.1 &  NE  \\
9   & 13.58  & $-$21:15.4     & 145.4    &   74.3 &  N \\
10 & 13.63  & $-$21:17.5     & 142.0    &   72.0 &  N \\
11 & 14.10  & $-$22:36.9     & 145.0    &   9.0   &  SW \\
15 & 12.98  & $-$22:08.4     & 129.4    &  29.0  &  NW \\
16 & 13.67  & $-$21:24.7     & 129.4    &   64.8 &   N \\
12 & 14.86  & $-$22:26.3     & 129.4    &   8.0   &  NE \\
13 & 14.66  & $-$22:27.8     & 129.4    &   4.7   &  NE \\
14 & 16.32  & $-$22:37.1     & 129.4    &  30.5  &  SEE \\
17 & 12.50  & $-$21:58.8     & 123.5    &   40.8 &  NW \\
18 & 12.66  & $-$22:34.9     & 123.5    &   26.1 & SWW \\                                                            
\hline
\end{tabular}
\end{center}
\vspace{0.1in}
Notes: 
(1) Seconds of Right Ascension.   
(2) Arc minutes and arc-seconds of Declination.  
(3) \Vlsr\  is the radial velocity in the Local Standard of Rest frame in which the OMC1 has  \Vlsr\ = 9 \kms .
(4) Projected distance in arc-seconds from the suspected ejection center located at 
J(2000) = 05$^h$35$^m$14.34$^s$, $-$5$^o$22'28.4".
(5) Direction from OMC1.
\end{table}

\clearpage

\end{document}